\documentclass[superscriptaddress,onecolumn,prb,preprintnumbers,amsmath,amssymb,showpacs]{revtex4}
\usepackage{lipsum} 
\usepackage[latin9]{inputenc}
\usepackage{graphicx}
\usepackage{dcolumn}
\usepackage{bbold}
\usepackage{bm}
\usepackage[usenames]{xcolor}

\usepackage[colorlinks,bookmarks=false,citecolor=blue,linkcolor=blue,urlcolor=blue,hyperfootnotes=true]{hyperref}

\makeatletter
\newsavebox\myboxA
\newsavebox\myboxB
\newlength\mylenA

\newcommand*\xoverline[2][0.75]{%
    \sbox{\myboxA}{$\m@th#2$}%
    \setbox\myboxB\null
    \ht\myboxB=\ht\myboxA%
    \dp\myboxB=\dp\myboxA%
    \wd\myboxB=#1\wd\myboxA
    \sbox\myboxB{$\m@th\overline{\copy\myboxB}$}
    \setlength\mylenA{\the\wd\myboxA}
    \addtolength\mylenA{-\the\wd\myboxB}%
    \ifdim\wd\myboxB<\wd\myboxA%
       \rlap{\hskip 0.5\mylenA\usebox\myboxB}{\usebox\myboxA}%
    \else
        \hskip -0.5\mylenA\rlap{\usebox\myboxA}{\hskip 0.5\mylenA\usebox\myboxB}%
    \fi}
\makeatother

\begin{document}

\title{Coexistence of $\Theta_{II}$-loop-current order with checkerboard $d$-wave CDW/PDW order
in a hot-spot model for cuprate superconductors}

\author{Vanuildo S. de Carvalho}
\affiliation{Instituto de Física, Universidade Federal de Goiás, 74.001-970,
Goiânia-GO, Brazil}
\author{Catherine Pépin}
\affiliation{IPhT, L'Orme des Merisiers, CEA-Saclay, 91191 Gif-sur-Yvette,
France}
\author{Hermann Freire}
\email{hermann\_freire@ufg.br}
\affiliation{Instituto de Física, Universidade Federal de Goiás, 74.001-970,
Goiânia-GO, Brazil}

\date{\today}
\begin{abstract}
We investigate the strong influence of the $\Theta_{II}$-loop-current order
on both unidirectional and bidirectional $d$-wave charge-density-wave/pair-density-wave (CDW/PDW) composite orders
along axial momenta $(\pm Q_0,0)$ and $(0,\pm Q_0)$ that emerge in an effective 
hot spot model departing from the three-band Emery model relevant to the phenomenology of the cuprate superconductors.
This study is motivated by the compelling evidence that the $\Theta_{II}$-loop-current order described by this model may explain groundbreaking experiments such as spin-polarized neutron scattering performed in these materials. Here, we demonstrate, within a saddle-point approximation, that the $\Theta_{II}$-loop-current order clearly coexists with
bidirectional (i.e. checkerboard) $d$-wave CDW and PDW orders along axial momenta, but is visibly detrimental to the unidirectional (i.e. stripe) case.
This result has potentially far-reaching implications for the physics of the cuprates and agrees well with very recent x-ray experiments on YBCO that indicate that at higher dopings the CDW order has indeed a tendency to be bidirectional. 
\end{abstract}

\pacs{74.20.Mn, 74.20.-z, 71.10.Hf}

\maketitle

\section{Introduction} 

There is now a sense of growing consensus in the high-$T_c$ superconductivity community that charge-density-wave (CDW) order \cite{Julien,Julien2,LeBoeuf,Ghiringhelli,Chang,Achkar,Hoffman,Yazdani} along axial momenta $(\pm Q_0,0)$ and $(0,\pm Q_0)$ (see Fig. \ref{Fermi_Surface_01}) with a $d$-wave form factor \cite{Comin,Fujita} is a universal phenomenon that takes place
inside the pseudogap phase in most underdoped cuprate superconductors. This realization provides crucial insights for the researchers in the field and potentially brings us one step closer to solving the long-standing problem concerning the nature of the pseudogap ``hidden" order
that favors the appearance of charge order in these materials. In this respect, a recent scanning tunneling microscope (STM) work with high-resolution by Hamidian \emph{et al.} \cite{Hamidian} has shed new light onto this fundamental question by demonstrating that the characteristic energy of the modulations of the CDW order is precisely the pseudogap energy. This result adds substantial support to the idea that the pseudogap phase is the primary order that sets in at a high temperature $T^*$, which generates a subsidiary charge order at $T_{CDW}<T^*$. The corresponding wavevectors describing such a charge order modulation connect approximately the so-called ``hot spots" (i.e. points in reciprocal space where the underlying Fermi arcs 
intersect the antiferromagnetic zone boundary) displayed in the pseudogap phase \cite{Comin2,SilvaNeto}.

Moreover, the phase diagram of the cuprates turned out to be amazingly richer than was previously thought. Spin-polarized neutron diffraction \cite{Fauque, Greven, Bourges_2, Sidis} indicates  a $\mathbf{q}=0$ intra-unit-cell time-reversal symmetry breaking occurring at the pseudogap temperature $T^*$. At a lower temperature $T_{Kerr}$, a tiny polar Kerr rotation has been reported as well  \cite{Kapitulnik,Kapitulnik_2}, indicating a breaking of both time-reversal symmetry and chirality. The proposal by Varma \cite{Varma} gives a natural explanation for the neutron scattering experiment in terms of 
a loop-current order (we shall focus here on the so-called $\Theta_{II}$-phase). Note that due to the lattice symmetries the presence of loop currents does not necessarily explain the polar Kerr effect: only loop currents breaking chiral symmetry rather than inversion symmetry, also called magneto-chiral states, produce a polar Kerr effect \cite{Yakovenko,Aji}. The description of the loop currents
requires, however, a minimal multiband model to begin with. For this reason, we shall depart from the standard Emery model \cite{Emery,Varma_2} that includes in addition to the usually considered copper $d_{x^2-y^2}$ orbital, 
also the oxygen $p_x$ and $p_y$ orbitals of the CuO$_2$ unit cell to account for the existence of intra-unit-cell
currents that show up in the $\Theta_{II}$-phase. Despite the strong appeal of such an order from the experimental viewpoint, it is important to emphasize here that this phase alone
is not expected to be the driving force of the pseudogap phase, since it does not break translational symmetry. Therefore,
it turns out to be difficult within this scenario to reproduce the result that the underlying Fermi surface (FS) partially gaps out in the antinodal regions, as seen experimentally.

Many theories for potential candidates of the pseudogap
``hidden" order were proposed in the literature
in order to try to resolve the above-mentioned discrepancies with
the experiments (see, e.g., \cite{PALee,Agterberg,Fradkin,Chowdhury,Wang_3,Atkinson,Pepin_4}). We will not detail those works here, but rather focus on the question whether the $\Theta_{II}$-loop-current
order suggested by Ref. \cite{Varma} as an explanation of the $\mathbf{q}=0$
neutron scattering experiment can coexist with the $d$-wave CDW order
observed in the underdoped regime inside the pseudogap phase. Our approach
will be rooted in the generally adopted spin-fluctuation scenario to the problem of high-$T_c$ superconductivity, in which it has been shown recently \cite{Efetov,Metlitski} that there is an emergent SU(2) symmetry relating a $d$-wave charge order along the diagonal direction and
the $d$-wave singlet superconducting order at the quantum critical point in the phase diagram.
In this context, another charge order (the experimentally observed $d$-wave CDW along axial momenta) also obtains an SU(2) partner in the form
of a $d$-wave pair-density-wave (PDW) state with the same wavevectors \cite{Pepin_2,Pepin_3,Wang_3,Freire,Freire_1}. The PDW order corresponds to a hypothesized superconducting order, which has a finite Cooper-pair center-of-mass momentum with some similarities to a Fulde-Ferrell-Larkin-Ovchinnikov (FFLO) state \cite{FFLO,FFLO_1}, but at zero magnetic field. The idea of a PDW state inside the pseudogap phase was put forward in
previous studies \cite{PALee,Agterberg,Fradkin}, and it was
shown to give a good explanation for the angle-resolved photoemission (ARPES) data in Bi2201\cite{PALee,Wang_3}.
In addition to this fact, another work in the literature \cite{Agterberg} has
explained that a PDW order may also give rise naturally to a secondary,
emergent phase that breaks both time-reversal and parity symmetry,
but preserves their product. As a result, this secondary
phase should exhibit the same symmetry properties of the $\Theta_{II}$-loop-current
phase, and therefore the possible existence of a PDW
order in the underdoped cuprates could be viewed as an alternative
explanation for the neutron scattering experiments \cite{Greven,Bourges_2}.

\section{Model and method} 

We start here with the three-band Emery model \cite{Emery,Varma_2}, which is given by $H=H_{0}+H_{int}$, i.e.
\vspace{-0.3cm}

\begin{align}
&\mathcal{H}_{0} = -t_{pd}\sum_{i,\sigma}\sum_{\nu}(\hat{d}_{i,\sigma}^{\dagger}\hat{p}_{i+\hat{\nu}/2,\sigma}+\text{H.c.})\nonumber \\
 & -t_{pp}\sum_{i,\sigma}\sum_{\langle\nu,\nu'\rangle}(\hat{p}_{i+\hat{\nu}/2,\sigma}^{\dagger}\hat{p}_{i+\hat{\nu}'/2,\sigma}+\text{H.c.})\nonumber \\
 & +(\varepsilon_{d}-\mu)\sum_{i,\sigma}\hat{n}_{i,\sigma}^{d}+\frac{1}{2}(\varepsilon_{p}-\mu)\sum_{i,\sigma}\sum_{\nu}\hat{n}_{i+\hat{\nu}/2,\sigma}^{p},\label{Eq_01}
\end{align} 

\begin{align}
& \mathcal{H}_{int} = U_{d}\sum_{i}\hat{n}_{i,\uparrow}^{d}\hat{n}_{i,\downarrow}^{d}+\frac{U_{p}}{2}\sum_{i,\nu}\hat{n}_{i+\hat{\nu}/2,\uparrow}^{p}\hat{n}_{i+\hat{\nu}/2,\downarrow}^{p}\nonumber \\
 & +V_{pd}\sum_{i,\nu}\sum_{\sigma,\sigma'}\hat{n}_{i,\sigma}^{d}\hat{n}_{i+\hat{\nu}/2,\sigma'}^{p},\label{Eq_02}
\end{align}

\noindent where $\hat{d}_{i,\sigma}^{\dagger}$, $\hat{d}_{i,\sigma}$, $\hat{p}_{i+\hat{\nu}/2,\sigma}^{\dagger}$,
and $\hat{p}_{i+\hat{\nu}/2,\sigma}$ stand for, respectively, the creation and annihilation
operators of fermions residing on the lattice site $i$ with spin projection $\sigma$
of the copper {[}Cu$(3d_{x^{2}-y^{2}})${]} orbital and the the creation and annihilation operators of fermions on the site $i+\hat{\nu}/2$
($\nu=x,y$) with spin projection $\sigma$ of the oxygen {[}O$(2p_{x})$
and O$(2p_{y})${]} orbitals in the CuO$_{2}$ unit cell. In addition, the parameters $t_{pd}$, $t_{pp}$, $U_{d}$, $U_{p}$ and $V_{pd}$ denote, respectively,
pair hoppings between $p_{x(y)}$ and $d$ orbitals and $p_x$ and $p_y$ orbitals, on-site interactions in the $d$ and $p_{x(y)}$ orbitals and nearest-neighbor interaction between the fermions on the copper  and oxygen orbitals. 
The quantities $\hat{n}_{i,\sigma}^{d}$ and $\hat{n}_{i+\hat{\nu}/2,\sigma}^{p}$
correspond to the fermionic number operators for particles located on the copper and oxygen orbitals and the parameters $\varepsilon_{d}$ and $\varepsilon_{p}$ 
are, respectively, the copper and oxygen orbital energies. Finally, $\mu$ is the chemical potential,
which changes the electronic density in the system.

\begin{figure}[t]
\begin{center}
\includegraphics[width=4in]{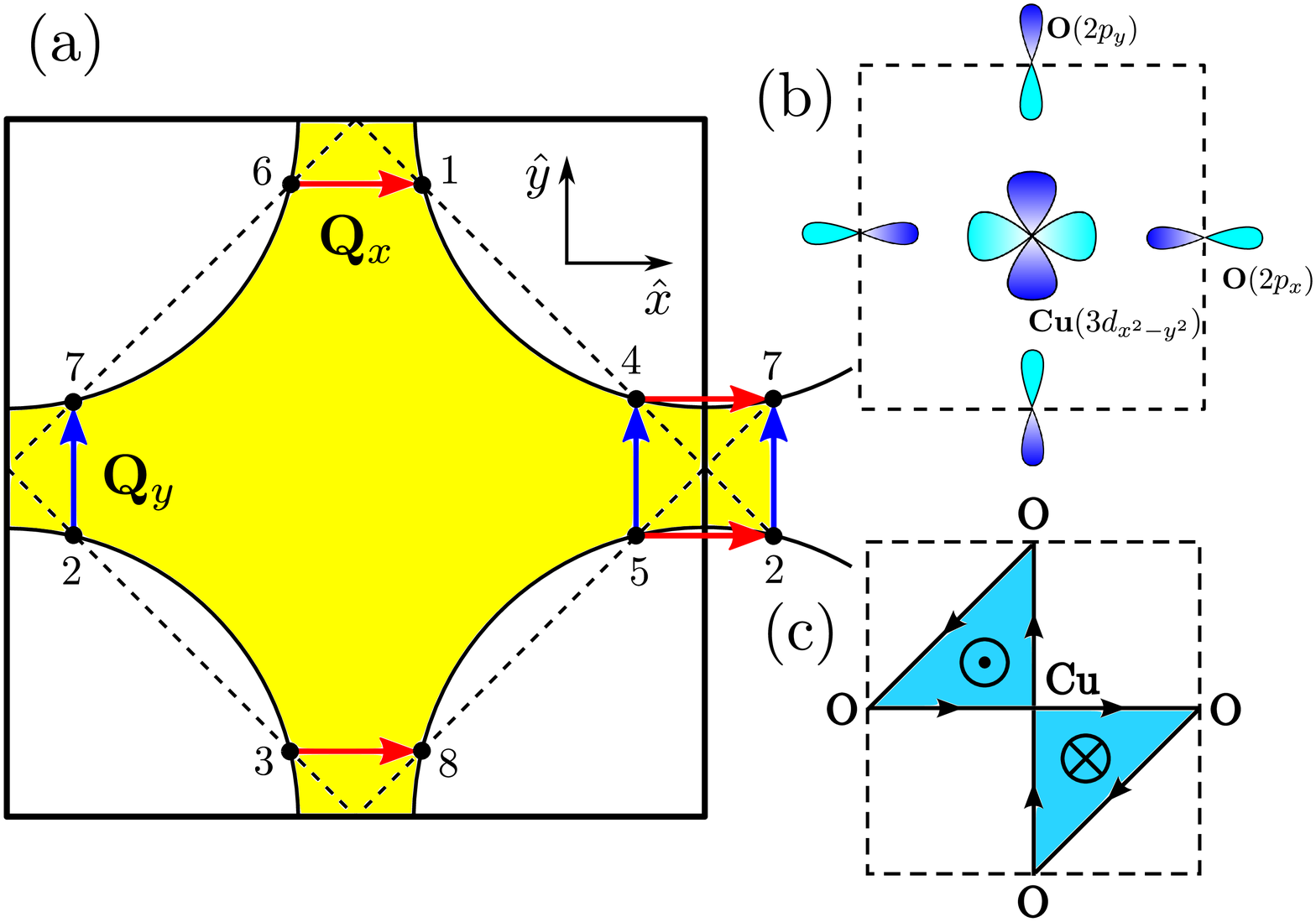}
\caption{(Color online) (a) Schematic representation of the underlying Fermi surface (enclosing the yellow area) that characterizes
the cuprates for experimentally relevant parameters. The vectors $\mathbf{Q}_{x}=(Q_0,0)$ and $\mathbf{Q}_{y}=(0, Q_0)$ represent the axial momenta. The numbered points refer to
the ``hot spots" that are defined as the intersection of the
Fermi surface with the antiferromagnetic zone boundary. The ``hot spot" with label 1 has a wavevector $\mathbf{k_1} = (K_{-},K_{+})$ in reciprocal space with the constraint $K_{-} + K_{+} =\pi$. The wavevectors of all the other ``hot spots" are obtained by simple rotation operations. (b) Structure of the copper [Cu$(3d_{x^2-y^2})$] and oxygen [O$(2p_x)$ and O$(2p_y)$] orbitals in the CuO$_2$ unit cell for the three-band Emery model. (c) Loop current pattern for
the $\Theta_{II}$-loop current phase. The symbols ($\odot$) and ($\otimes$) stand for
the orientation of the local magnetic moments generated by the loop currents. }\label{Fermi_Surface_01}
\end{center}
\end{figure}

Here, we implement the standard method of decoupling the $U_d$ term in the magnetic channel to derive the spin-fermion model \cite{Chubukov,Chubukov_1}. Then, by expressing the $V_{pd}$ term as a sum of current operators, we also decouple this interaction to incorporate the definition of the $\Theta_{II}$-loop-current order parameter \cite{Varma,Fischer} that generates the current pattern shown in Fig. \ref{Fermi_Surface_01}. For physical choices of the parameters in the model, the lowest energy band will lead to the experimentally relevant Fermi surface also depicted schematically in Fig. \ref{Fermi_Surface_01}. It is crucial to emphasize here that the most important contribution in this model will originate from the ``hot spots" mentioned above. Therefore, we shall restrict the present analysis to the vicinity of such ``hot spots" and linearize the excitation spectrum of the three-band model around those points. In this way, we define the 16-component spinors $d=(d_{1},\ldots,d_{8})^{\text{T}}$ and $p_{x(y)
}=(p_{x(y)1},\ldots,p_{x(y)8})^{\text{T}}$, where $d_{i}=(d_{i,\uparrow}, d_{i,\downarrow})^{\text{T}}_{\sigma}$ and $p_{x(y)i}=(p_{x(y)i,\uparrow}, p_{x(y)i,\downarrow})^{\text{T}}_{\sigma}$ are also spinors in the spin space $\sigma$ which act in the neighborhood of each hot spot labeled by $i$. As a result, we obtain the following effective action describing the system
\begin{widetext} 
\vspace{-0.3cm}

\begin{align}\label{Eq_09}
 & \mathcal{S}[p_{x},p_{y},d,\vec{\phi};n_{p},R_{II}]=\mathcal{S}_{0}[p_{x},p_{y},d]+\mathcal{S}_{int}^{(1)}[d,\vec{\phi}]+\mathcal{S}_{int}^{(2)}[p_{x},p_{y},d;n_{p},R_{II}]\nonumber \\
 & =\int\begin{pmatrix}p_{x}^{\dagger}(X), & p_{y}^{\dagger}(X), & d^{\dagger}(X)\end{pmatrix}\begin{pmatrix}\partial_{\tau}+\xi_{p} & \hat{\Gamma}_{1}+\hat{\Gamma}_{2}(-i\nabla) & \hat{\Gamma}_{1x}-\hat{\Gamma}_{2x}i\partial_{x}\\
\hat{\Gamma}_{1}+\hat{\Gamma}_{2}(-i\nabla) & \partial_{\tau}+\xi_{p} & \Gamma_{1y}-\Gamma_{2y}i\partial_{y}\\
\hat{\Gamma}_{1x}^{\dagger}-\hat{\Gamma}_{2x}^{\dagger}i\partial_{x} & \hat{\Gamma}_{1y}^{\dagger}-\hat{\Gamma}_{2y}^{\dagger}i\partial_{y} & \partial_{\tau}+\xi_{d}
\end{pmatrix}\begin{pmatrix}p_{x}(X)\\
p_{y}(X)\\
d(X)
\end{pmatrix}dX\nonumber \\
 & +\frac{1}{2}\int\left[\frac{1}{v_{s}^{2}}(\partial_{\tau}\vec{\phi})^{2}+(\nabla\vec{\phi})^{2}+m_a\vec{\phi}^{2}\right]dX+\lambda\int\left[d^{\dagger}(X)\Sigma_{1}\vec{\phi}(X)\vec{\sigma}d(X)\right]dX +\int\left(\frac{R_{II}^{2}}{V_{pd}}-\frac{n_{p}^{2}}{8}U_{p}\right)dX,
\end{align}
\end{widetext} 

\noindent where the bosonic field $\vec{\phi}_{i}=(\phi_{i}^{x},\phi_{i}^{y},\phi_{i}^{z})$
is the spin-density-wave (SDW) order parameter at the antiferromagnetic
wave vector $\mathbf{Q}=(\pi,\pi)$, $v_{s}$ is the spin-wave velocity,
$m_a$ is the spin-wave bosonic mass which vanishes at the quantum
critical point of the theory, and $\lambda$ is the spin-fermion coupling constant \cite{Chubukov,Chubukov_1}. $R_{II}$ is the standard $\Theta_{II}$-loop-current order parameter \cite{Varma,Fischer}, which is defined for completeness in the Appendix A. The $\sigma^{a}$ $(a=x,y,z)$
are the usual Pauli matrices. Besides, $\xi_{p}\equiv\varepsilon_{p}+\frac{n_{p}}{4}U_{p}-\mu$, $\xi_{d}\equiv\varepsilon_{d}-\mu$, and the variable $X=(\tau,\textbf{r})$
includes both time and space coordinates. The matrix $\Sigma_1$ refers to the $x$-component of the Pauli matrices defined in a pseudospin space \cite{Efetov}.
The other matrices are diagonal in an enlarged $\Sigma\otimes\Lambda\otimes L$ pseudospin space \cite{Efetov} and
are given by $\hat{\Gamma}_{1}=-2t_{pp}\cos\delta\ \mathbb{1}_{\Sigma}\otimes\mathbb{1}_{\Lambda}\otimes\mathbb{1}_{L}$,
$\hat{\Gamma}_{2}=t_{pp}(\sin\delta\Lambda_{3}\otimes L_{3}-\Sigma_{3}\otimes\Lambda_{3})i\partial_{x} -t_{pp}(\sin\delta\Lambda_{3}+\Sigma_{3}\otimes\Lambda_{3}\otimes L_{3})i\partial_{y}$, 
$\hat{\Gamma}_{1x}=\gamma_{1}e^{-i\varphi\Lambda_{3}\otimes L_{3}}+\gamma_{2}e^{i\theta\Lambda_{3}\otimes L_{3}}\Sigma_{3}\otimes L_{3}$,
$\hat{\Gamma}_{2x}=-\frac{1}{2}\gamma_{1}e^{-i\varphi\Lambda_{3}\otimes L_{3}}\Sigma_{3}\otimes\Lambda_{3}+\frac{1}{2}\gamma_{2}e^{i\theta\Lambda_{3}\otimes L_{3}}\Lambda_{3}\otimes L_{3}$, 
$\hat{\Gamma}_{1y}=\gamma_{1}e^{i\varphi\Lambda_{3}}-\gamma_{2}e^{-i\theta\Lambda_{3}}\Sigma_{3}\otimes L_{3}$, and
$\hat{\Gamma}_{2y}=-\frac{1}{2}\gamma_{1}e^{i\varphi\Lambda_{3}}\Sigma_{3}\otimes\Lambda_{3}\otimes L_{3}+\frac{1}{2}\gamma_{2}e^{-i\theta\Lambda_{3}}\Lambda_{3}$,
where $\delta=(K_{+}-K_{-})/2$ (see Fig. \ref{Fermi_Surface_01}) and $\mathbb{1}_{\Sigma}$, $\mathbb{1}_{\Lambda}$,
and $\mathbb{1}_{L}$ denote, respectively, the identity matrices in
the $\Sigma$, $\Lambda$, and $L$ pseudospin spaces. The parameters
$\varphi$, $\theta$, $\gamma_{1}$, and $\gamma_{2}$ are defined
as $\tan\varphi=\frac{R_{II}}{2t_{pd}}\tan(\frac{\delta}{2})$,
$\tan\theta=\frac{R_{II}}{2t_{pd}}\cot(\frac{\delta}{2})$,
$\gamma_{1}=\sqrt{2t_{pd}^{2}\cos^{2}(\frac{\delta}{2})+\frac{R_{II}^{2}}{2}\sin^{2}(\frac{\delta}{2})}$, and
$\gamma_{2}=\sqrt{2t_{pd}^{2}\sin^{2}(\frac{\delta}{2})+\frac{R_{II}^{2}}{2}\cos^{2}(\frac{\delta}{2})}$.

To compute the thermodynamical properties of the present model,
we first integrate out the bosonic field in the functional
integral to derive an effective quartic interaction for the fermionic fields. We defer the explanation of this procedure to the Appendices A and B. After that, the next step is to decouple the
fermionic quartic term by using a $d$-wave composite order parameter $\hat{M}_{\alpha}(X,X')$ for either unidirectional ($\alpha$=CDW-1/PDW-1) or bidirectional ($\alpha$=CDW-2/PDW-2) intertwined CDW/PDW phase along axial momenta $(\pm Q_0,0)$ and $(0,\pm Q_0)$. The order parameter $\hat{M}_{\alpha}(X,X')$ for the CDW/PDW composite order can be written as

\vspace{-0.4cm}

\begin{align}
 & \hat{M}_{\alpha}(X,X')=-b_{\alpha}(X,X')\biggl[\Sigma_{3}\otimes\Lambda_{3}\otimes\begin{pmatrix}0 & \hat{u}_{\tau}\\
-\hat{u}_{\tau}^{\dagger} & 0
\end{pmatrix}_{L}\nonumber\\
 & +\delta_{\alpha,\text{CDW-2/PDW-2}}\Sigma_{3}\otimes L_{1}\otimes\begin{pmatrix}0 & \hat{u}_{\tau}\\
-\hat{u}_{\tau}^{\dagger} & 0
\end{pmatrix}_{\Lambda}\biggr],\label{Eq_18}\\
 & \text{with}\hspace{0.5cm}\hat{u}_{\tau}=\begin{pmatrix}\Delta_{\text{CDW}} & \Delta_{\text{PDW}}\\
-\Delta_{\text{PDW}}^{*} & \Delta_{\text{CDW}}^{*}
\end{pmatrix}_{\tau}.\label{Eq_19}
\end{align}
Here, $\Delta_{\text{CDW}}$ and $\Delta_{\text{PDW}}$ denote, respectively, the $d$-wave CDW and PDW components
of the order parameter defined above. The matrices
$\hat{u}_{\tau}$ belong to the SU(2) group, such that $|\Delta_{\text{CDW}}|^{2}+|\Delta_{\text{PDW}}|^{2}=1$ involving both the PDW and CDW sectors. As a result, the mean-field equation for $b_{\alpha}(\varepsilon_{n},\mathbf{k})$ in terms of $\mathcal{D}^{\ell m}_{\alpha}(i\varepsilon_{n},\mathbf{k})$
yields

\vspace{-0.4cm}

\begin{align}\label{Eq_31}
b_{\alpha}(\varepsilon_{n},\mathbf{k}) & =\frac{3\lambda^{2}T}{16\zeta}\sum\limits _{\ell,m=1}^{2}\sum\limits _{\varepsilon'_{n}}^ {}\int\dfrac{D_{eff}(\varepsilon_{n}-\varepsilon'_{n},\mathbf{k}-\mathbf{k}')}{\mathcal{D}^{\ell m}_{\alpha}(i\varepsilon'_{n},\mathbf{k}')}\nonumber \\
 & \times\dfrac{\partial\mathcal{D}^{\ell m}_{\alpha}(i\varepsilon'_{n},\mathbf{k}')}{\partial b_{\alpha}(\varepsilon'_{n},\mathbf{k}')}\frac{d\mathbf{k}'}{(2\pi)^{2}},
\end{align}

\noindent where $D_{eff}(\omega,\mathbf{k})=(\gamma|\omega|+|\mathbf{k}|^{2}+m_{a})^{-1}$
and $\gamma$ is the Landau damping term. In what follows, we set $\zeta=1$ for unidirectional order and $\zeta=2$ for bidirectional order. Moreover, to
calculate self-consistently the mean-field loop-current order parameter $R_{II}$, we proceed to minimize 
the free energy of the model with respect to $R_{II}$. Thus, we obtain that the mean-field
equation for this order parameter reads as

\vspace{-0.4cm}

\begin{equation}\label{Eq_33}
R_{II}=\frac{V_{pd}T}{2}\sum\limits _{\ell,m=1}^{2}\sum_{\varepsilon_{n}}\int\frac{1}{\mathcal{D}^{\ell m}_{\alpha}(i\varepsilon_{n},\mathbf{k})}\frac{\partial\mathcal{D}^{\ell m}_{\alpha}(i\varepsilon_{n},\mathbf{k})}{\partial R_{II}}\frac{d\mathbf{k}}{(2\pi)^{2}}.
\end{equation}

\section{Numerical results}

To study the effect of the $\Theta_{II}$-loop-current 
on both $d$-wave CDW and PDW orders, we solve numerically
the mean-field equations for $R_{II}$ and $b_{\alpha}$. In order to keep this computation tractable, we discretize the Brillouin zone using a grid of $320\times320$ points around the ``hot spots". We have checked that the solution for the the gap equations does not change qualitatively as the number of points on the grid is increased. In addition, we neglect the space-time distribution of the order parameter $b_{\alpha}(\varepsilon_{n},\mathbf{k})$ by eliminating its dependence on frequency and momentum. As a consequence, this enables us to evaluate the Matsubara sums
that appear in the mean-field equations in an analytical way. We perform this computation as a function of the spin-fermion coupling $\lambda$, such that the nearest-neighbor interaction $V_{pd}$ is of the same order of magnitude as the coupling $\lambda$ (it is reasonable to assume that this physical regime should apply to some high-$T_c$ compounds). By contrast, we would like to point out that outside this regime one cannot find the coexistence of LC and CDW/PDW orders, which will be described in more detail below. Hence, we set throughout this work the physically-motivated
parameters in the effective model as follows: $t_{pd}=1$, $t_{pp}=0.5$, $U_{p}=3$, $\varepsilon_{d}-\varepsilon_{p}=3$, $\gamma=10^{-5}$, $m_a=10^{-3}$ and $V_{pd}=\lambda$. 
The occupation number on the oxygen orbital is given by $n_{p}\approx 0.3$ and the occupation number on the copper orbital is given by $n_d+2n_p=1+x$ (where $x$ is the hole doping parameter) and $m_a\propto (x-x_c)$, with $x_c$ being the critical doping associated with the putative quantum critical point close to optimal doping.

\begin{figure*}[t]
\includegraphics[width=6.5in]{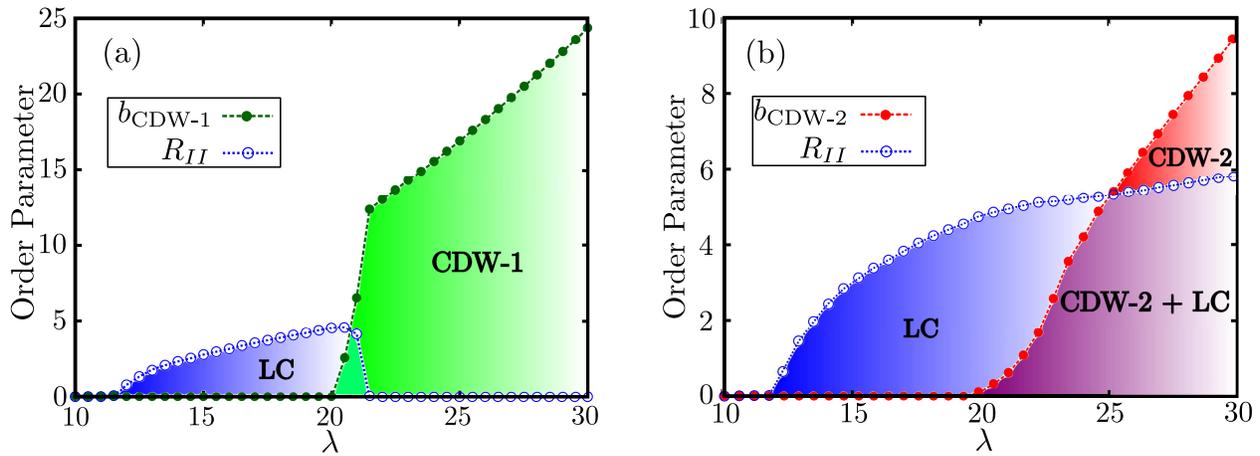}
\caption{(Color online) (a) Mean-field values of the $\Theta_{II}$-loop-current order parameter $R_{II}$ and unidirectional $d$-wave CDW (with label CDW-1) as a function
of the spin-fermion coupling constant $\lambda$ in the limit of zero
temperature for the nearest-neighbor interaction $V_{pd}$, such that $V_{pd}=\lambda$. (b) Mean-field values of the $\Theta_{II}$-loop-current (LC) order parameter $R_{II}$ and bidirectional $d$-wave CDW (with label CDW-2) as a function
of the spin-fermion coupling constant $\lambda$ in the limit of zero
temperature for nearest-neighbor interaction $V_{pd}$, such that $V_{pd}=\lambda$.
Both plots (a) and (b) were obtained by performing numerical integration in momentum
space of the self-consistency equations given by Eqs. \eqref{Eq_31}
and \eqref{Eq_33} with a mesh of $320\times320$ points in the Brillouin
zone. Here, $m_{a}=10^{-3}$, $\gamma=10^{-5}$ and the other parameters
are set to $t_{pd}=1$, $t_{pp}=0.5$, $U_{p}=3$, and $\varepsilon_{d}-\varepsilon_{p}=3$.
The occupation number on the oxygen orbital is given by $n_{p}\approx 0.3$ and the occupation number on the copper orbital is given by $n_d+2n_p=1+x$, where $x$ is the hole doping parameter and $m_a\propto (x-x_c)$ with $x_c$ being the critical doping associated with the quantum critical point.
}\label{1D_CDW_02}
\end{figure*}

In view of the fact that some recent experimental works reported the possible emergence of a unidirectional (i.e. stripe) charge order taking place at low doping in the cuprates \cite{Damascelli,Davis}, we first consider this case here. By solving numerically Eqs. (\ref{Eq_31}) and (\ref{Eq_33}), we observe from Fig. \ref{1D_CDW_02} (a) that as the spin-fermion interaction $\lambda$ is increased, the stripe $d$-wave CDW order parameter clearly grows from zero to finite values. By contrast, the $R_{II}$ order parameter displays the opposite effect, namely, as the CDW achieves finite values, the $R_{II}$ order becomes clearly suppressed. This result suggests that the unidirectional $d$-wave CDW order with a modulation along axial momenta is seemingly detrimental to the $\Theta_{II}$-loop-current order (and vice versa), and the general tendency observed from our data is that these two orders tend not to coexist in the present model. We point out that this conclusion is reminiscent of a different study made previously by us in \cite{de_Carvalho} 
with 
collaborators, in which we 
determined numerically that the $\Theta_{II}$-loop-current order strongly competes with the $d$-wave quadrupolar density wave with a modulation along the diagonal directions, which was widely discussed in many theoretical works (see, e.g, \cite{Metlitski,Efetov,Pepin_2,Pepin_3,Freire,Freire_1}). As a result, we offered this scenario as a possible explanation as to why such a quadrupolar density wave order with diagonal modulation is never observed experimentally, e.g., in x-ray \cite{Ghiringhelli,Chang,Achkar} and STM experiments \cite{Yazdani,Comin}. 

\begin{figure*}[t]
\includegraphics[width=6.5in]{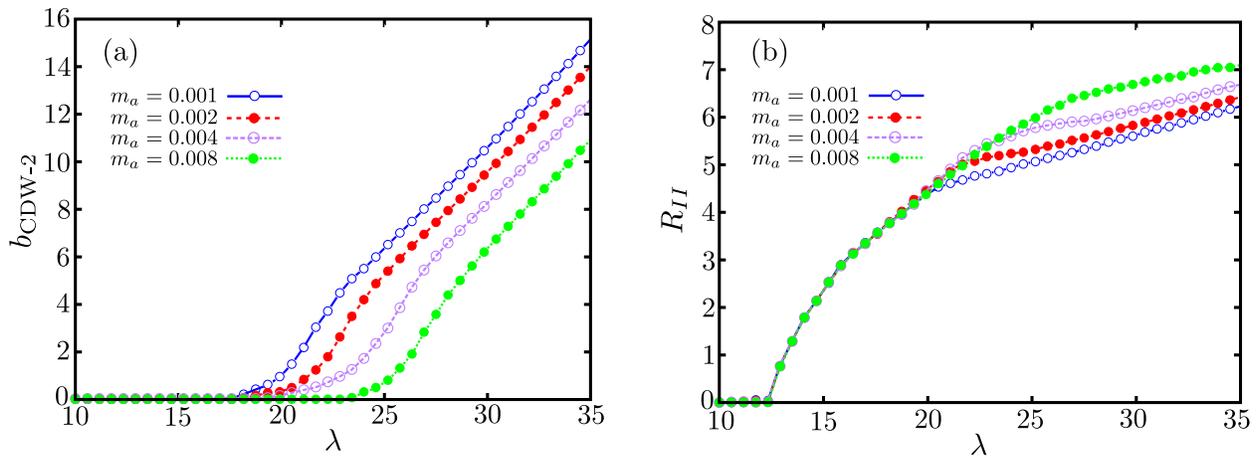}
\caption{(Color online) (a) Mean-field values of the bidirectional $d$-wave CDW order parameter as a function
of the spin-fermion coupling constant $\lambda$ in the limit of zero
temperature for the nearest-neighbor interaction $V_{pd}$ such that $V_{pd}=\lambda$, for several choices of spin-wave bosonic mass $m_a$ that controls the hole doping in the model. (b) Mean-field values of the $\Theta_{II}$-loop-current (LC) order parameter as a function
of the spin-fermion coupling constant $\lambda$ in the limit of zero
temperature for the nearest-neighbor interaction $V_{pd}$, such that $V_{pd}=\lambda$ for several choices of spin-wave bosonic mass $m_a$.
Both plots (a) and (b) were obtained by performing numerical integration in momentum
space of the self-consistency equations given by Eqs. \eqref{Eq_31}
and \eqref{Eq_33} with a mesh of $320\times320$ points in the Brillouin
zone. Besides, $\gamma=10^{-5}$, $t_{pd}=1$, $t_{pp}=0.5$, $U_{p}=3$, and $\varepsilon_{d}-\varepsilon_{p}=3$.
The occupation number on the oxygen orbital is given by $n_{p}\approx 0.3$ and the occupation number on the copper orbital is given by $n_d+2n_p=1+x$, where $x$ is the hole doping parameter and $m_a\propto (x-x_c)$ with $x_c$ being the critical doping associated with the quantum critical point.
}\label{2D_CDW_02}
\end{figure*}

We now move on to the analysis of the $d$-wave CDW bidirectional (i.e. checkerboard) case. The corresponding results are shown in Fig. \ref{1D_CDW_02} (b). Similar to the previous situation, we note that, as the interaction parameter $\lambda$ increases, such a bidirectional CDW also becomes finite. However, in marked contrast to the previous case, this order does not suppress the $\Theta_{II}$-loop-current order. Indeed, we observe from our data that there is clearly a region where these two orders coexist, implying that they do not compete with each order, at least at mean-field level. This result is consistent with neutron diffraction experiments \cite{Greven,Bourges_2} (regarding the evidence of time-reversal symmetry-breaking), and also with
very recent x-ray data in YBCO \cite{LeTacon} that complements this scenario by posing at higher dopings the subsidiary $d$-wave charge order that emerges at lower temperatures turns out to be indeed of bidirectional nature. In this respect, a theoretical work by Wang and Chubukov \cite{Chubukov_2} appeared recently using a Ginzburg-Landau analysis for a different hot spot model and they concluded that initially the CDW order is in fact bidirectional but changes to unidirectional inside the CDW dome at low enough doping, which would explain apparently conflicting experimental data published in the literature \cite{Damascelli,Davis,LeTacon}. By contrast, at large enough doping, the CDW order turns out to be always bidirectional according to their scenario. Our results presented here in this work agree qualitatively with their conclusions.  

To further confirm the qualitative comparison of the present results with the experimental data in the cuprate superconductors, we next plot the values of both the bidirectional $d$-wave CDW and the $\Theta_{II}$-loop current order parameters at $T=0$ as a function of the coupling $\lambda$ for the regime such that the nearest-neighbor interaction $V_{pd}$ is given by $V_{pd}=\lambda$. We perform this computation for some choices of the spin-wave bosonic mass $m_a$, which controls the hole doping $x$ in the present effective model. The numerical results are shown in Figs. \ref{2D_CDW_02} (a) and (b). In Fig. \ref{2D_CDW_02} (a), we observe that the main effect of moving away from the quantum critical point of the present theory close to optimal doping corresponds essentially to shift the critical spin-wave coupling associated with the emergence of the bidirectional $d$-wave CDW phase in the model towards larger values. This scenario would in principle explain qualitatively why it is more likely to observe the bidirectional $d$-wave CDW short-range order in slightly overdoped cuprate samples, as seen experimentally, e.g., in Ref. \cite{LeTacon}. On the other hand, we note in Fig. \ref{2D_CDW_02} (b) that a similar variation in the carrier concentration of the present model apparently has no strong effect on the critical coupling associated with the appearance of the $\Theta_{II}$-loop-current order parameter. This indicates that this latter phase is seemingly more robust in the present model (at least at mean-field level), and no fine-tuning of the doping parameter is necessary for the emergence of the $\Theta_{II}$-loop-current order. This also agrees qualitatively with experimental results \cite{Greven,Bourges_2}.

Lastly, another important aspect of the present model that it is worth commenting on concerns the emergent SU(2) pseudospin symmetry \cite{Metlitski,Efetov} in the effective hot spot model, which relates a $d$-wave bidirectional CDW order to an yet undetected $d$-wave bidirectional PDW order \cite{Freire,Freire_1} (the same symmetry also holds for unidirectional $d$-wave CDW and PDW orders \cite{Pepin_2,Pepin_3,Wang_2}). 
According to our present result, this symmetry implies that the $R_{II}$ order parameter can also coexist with a checkerboard PDW order in the model for the same interaction regime as displayed in Fig. \ref{2D_CDW_02}. 
Note that this is probably not the case for unidirectional PDW. In this second situation, such a stripe PDW is expected to be damaging to the $\Theta_{II}$-loop-current order (and vice versa). For this reason, these two latter phases will tend not to coexist in the model. 

\section{Conclusions} 

We have analyzed the effect that the $\Theta_{II}$-loop-current order has
on stripe and checkerboard $d$-wave CDW and PDW orders
along axial momenta in a 
hot spot model, which is derived from the standard three-band Emery model.
We have shown, within a saddle-point approximation, that the $\Theta_{II}$-loop-current order clearly coexists with
a checkerboard $d$-wave CDW order along axial momenta.
We have also pointed out that the $\Theta_{II}$-loop-current order is likely to be harmful to stripe $d$-wave CDW order in the model.
Quite amazingly, this result turns out to be consistent with very recent x-ray experiments \cite{LeTacon} on YBCO that provided solid evidence that at higher doping concentrations the short-range $d$-wave CDW order that manifests itself in this compound has indeed a tendency to be of bidirectional nature. Due to an emergent SU(2) pseudospin symmetry \cite{Metlitski,Efetov,Freire,Freire_1,Pepin_2,Pepin_3,Wang_2}  that exists in the low-energy effective hot spot model, we expect that an accompanying bidirectional $d$-wave PDW order may also appear in this system. Therefore, we hope that our present theoretical result will stimulate further experimental studies to continue looking for fingerprints of such a novel $d$-wave checkerboard PDW order that is likely to be present in some cuprate materials.

\acknowledgments

The authors wish to acknowledge CAPES (V.S. de C.) and CNPq (H.F.) for financial support. C.P. has received financial support from LabEx PALM (Grant No. ANR-10-LABX-0039-PALM), ANR project UNESCOS ANR-14-CE05-0007, and the COFECUB Grant No. Ph743-12. We also thank Yvan Sidis and Ph. Bourges for useful discussions.

\appendix

\section{Mean-field equations}

Here, we explain some technical aspects of the effective field theory describing the interplay between the so-called $\Theta_{II}$-loop-current (LC) order, charge-density-wave (CDW) and pair-density-wave (PDW) order in a two-dimensional hot spot model for the cuprate superconductors. The approach for studying these orders consists of the derivation of the so-called spin-fermion (SF) model departing from a three-band model defined in the CuO$_{2}$ unit cell and then decoupling the Hubbard interaction between the O and Cu orbitals in operators that generate the current pattern observed in neutron scattering experiments. Integrating out the bosonic modes and subsequently the fermionic fields of the excitations around the hot spots, we determine that the functional free energy of the model in space-time coordinates $X=(\tau,\mathbf{r})$ is given by
\begin{align}\label{S01}
 \frac{\mathcal{F}_{\alpha}[T,n_{p},R_{II},\hat{M}_{\alpha}]}{T}&=-\int\text{Tr}\ln[G^{-1}_{\alpha}(X,X')]dXdX'+\frac{1}{2}\int J^{-1}(X-X')\text{Tr}[\hat{M}_{\alpha}(X,X')\Sigma_{1}\hat{M}_{\alpha}(X',X)\Sigma_{1}]dXdX'\nonumber \\
 & +\int\left(\frac{R_{II}^{2}}{V_{pd}}-\frac{n_{p}^{2}}{8}U_{p}\right)dX,
\end{align}
where $J(X-X')=3\lambda^{2}D_{eff}(X-X')$ is the effective propagator for the paramagnons, $\hat{M}_{\alpha}(X,X')$ is the order parameter for either unidirectional ($\alpha$=CDW-1/PDW-1) or bidirectional ($\alpha$=CDW-2/PDW-2) intertwined CDW/PDW order with a $d$-wave form factor, and $R_{II}=-iV_{pd}\sum_{\sigma}\langle\mathcal{A}_{i,\sigma}\rangle$ is the order parameter associated with the LC phase with the operator $\mathcal{A}_{i,\sigma}$ being given by
\begin{equation}
\mathcal{A}_{i,\sigma}=\frac{i}{2}[(\hat{d}_{i,\sigma}^{\dagger}\hat{p}_{i+\hat{x}/2,\sigma}-\hat{d}_{i,\sigma}^{\dagger}\hat{p}_{i-\hat{x}/2,\sigma})+(\hat{d}_{i,\sigma}^{\dagger}\hat{p}_{i+\hat{y}/2,\sigma}-\hat{d}_{i,\sigma}^{\dagger}\hat{p}_{i-\hat{y}/2,\sigma})].
\end{equation}
The matrix $G^{-1}_{\alpha}(X,X')$ is the Fourier transform of $G^{-1}_{\alpha}(i\varepsilon_{n},\mathbf{k})$, which is given by
\begin{equation}\label{S02}
\begin{pmatrix}-i\varepsilon_{n}+\xi_{p}\tau_{3} & \hat{\Gamma}_{1}\tau_{3}-\hat{\Gamma}_{2}(\mathbf{k}) & \hat{\Gamma}_{1x}\tau_{3}-\hat{\Gamma}_{2x}k_{x}\\
\hat{\Gamma}_{1}^{\dagger}\tau_{3}-\hat{\Gamma}_{2}^{\dagger}(\mathbf{k}) & -i\varepsilon_{n}+\xi_{p}\tau_{3} & \hat{\Gamma}_{1y}\tau_{3}-\hat{\Gamma}_{2y}k_{y}\\
\hat{\Gamma}_{1x}^{\dagger}\tau_{3}-\hat{\Gamma}_{2x}^{\dagger}k_{x} & \hat{\Gamma}_{1y}^{\dagger}\tau_{3}-\hat{\Gamma}_{2y}^{\dagger}k_{y} & -i\varepsilon_{n}+\xi_{d}\tau_{3}-i\hat{M}_{\alpha}(\varepsilon_{n},\mathbf{k})
\end{pmatrix},
\end{equation}
with $\hat{\Gamma}_{1}$, $\hat{\Gamma}_{2}$, $\hat{\Gamma}_{1x(y)}$, and $\hat{\Gamma}_{2x(y)}$ being matrices defined in pseudospin space $\Sigma\otimes\Lambda\otimes L$ that we have written explicitly in the main text. For this reason, we shall not repeat them here.

The order parameter for unidirectional (i.e. stripe) $d$-wave CDW/PDW in the case of a linearized spectrum around the hot spots possesses an SU(2) symmetry relating both phases. Here, we choose the wavevector of such a stripe phase to be $\mathbf{Q}_{x}=(Q_0,0)$, where $Q_0$ measures the distance between two neighboring hot spots along the $x$-axis. Therefore, following the approach of Refs. \cite{Pepin_2,Pepin_3,Efetov,de_Carvalho}, its order parameter can be written as
\begin{align}
 \hat{M}_{\text{CDW-1/PDW-1}}(i\varepsilon_{n},\mathbf{k})&=-b_{\text{CDW-1/PDW-1}}(i\varepsilon_{n},\mathbf{k})\Sigma_{3}\otimes\Lambda_{3}\otimes\begin{pmatrix}0 & \hat{u}_{\tau}\\
-\hat{u}_{\tau}^{\dagger} & 0
\end{pmatrix}_{L},\label{S13}\\
 \text{with} \hspace{0.4cm}&\hat{u}_{\tau}=\begin{pmatrix}\Delta_{\text{CDW}} & \Delta_{\text{PDW}}\\
-\Delta_{\text{PDW}}^{*} & \Delta_{\text{CDW}}^{*}
\end{pmatrix}_{\tau}.\label{S14}
\end{align}
By definition, $\hat{u}_{\tau}$ is a matrix of the SU(2) group and the constraint $|\Delta_{\text{CDW}}|^{2}+|\Delta_{\text{PDW}}|^{2}=1$ always holds. For the case of a bidirectional (or checkerboard) $d$-wave CDW/PDW phase, its modulations occur for both wavevectors $\mathbf{Q}_{x}=(Q_0,0)$ and $\mathbf{Q}_{y}=(0,Q_0)$ connecting pairs of hot spots in the $x$- and $y$-directions of the Brillouin zone. As a result, the order parameter for this phase assumes the form 
\begin{equation}\label{S15}
 \hat{M}_{\text{CDW-2/PDW-2}}(i\varepsilon_{n},\mathbf{k})=-b_{\text{CDW-2/PDW-2}}(i\varepsilon_{n},\mathbf{k})\biggl[\Sigma_{3}\otimes\Lambda_{3}\otimes\begin{pmatrix}0 & \hat{u}_{\tau}\\
-\hat{u}_{\tau}^{\dagger} & 0
\end{pmatrix}_{L}+\Sigma_{3}\otimes L_{1}\otimes\begin{pmatrix}0 & \hat{u}_{\tau}\\
-\hat{u}_{\tau}^{\dagger} & 0
\end{pmatrix}_{\Lambda}\biggr].
\end{equation}

The mean-field equations for both $d$-wave CDW/PDW composite order and LC phase are obtained from Eq. \eqref{S01} by demanding, respectively, that $\delta\mathcal{F}_{\alpha}/\delta b_{\alpha}=0$ and $\delta\mathcal{F}_{\alpha}/\delta R_{II}=0$. By Fourier transforming those equations to frequency-momentum space, they finally become
\begin{align}
 b_{\alpha}(\varepsilon_{n},\mathbf{k})&=\frac{3\lambda^{2}T}{16\zeta}\sum_{\varepsilon'_{n}}\int\frac{D_{eff}(\varepsilon_{n}-\varepsilon'_{n},\mathbf{k}-\mathbf{k}')}{\det[G^{-1}_{\alpha}(i\varepsilon'_{n},\mathbf{k}')]}\frac{\partial\det[G^{-1}_{\alpha}(i\varepsilon'_{n},\mathbf{k}')]}{\partial b_{\alpha}(\varepsilon'_{n},\mathbf{k}')}\frac{d\mathbf{k}'}{(2\pi)^{2}},\label{S16}\\
 R_{II}&=\frac{V_{pd}T}{2}\sum_{\varepsilon_{n}}\int\frac{1}{\det[G^{-1}_{\alpha}(i\varepsilon_{n},\mathbf{k})]}\frac{\partial\det[G^{-1}_{\alpha}(i\varepsilon_{n},\mathbf{k})]}{\partial R_{II}}\frac{d\mathbf{k}}{(2\pi)^{2}},\label{S17}
\end{align}
where we set $\zeta=1$ for stripe $d$-wave CDW/PDW order and $\zeta=2$ for checkerboard $d$-wave CDW/PDW order.

\section{Evaluation of the determinant of the matrix $G^{-1}_{\alpha}(i\varepsilon_{n},\mathbf{k})$}

We now turn our attention to the evaluation of $\det[G^{-1}_{\alpha}(i\varepsilon_{n},\mathbf{k})]$. In order to do that, we will need to use the set of determinant formulas
\begin{align}
 &\det\begin{pmatrix} \hat{A} & \hat{B} \\ \hat{C} & \hat{D} \end{pmatrix}=\det(\hat{A})\det(\hat{D}-\hat{C}\hat{A}^{-1}\hat{B}),\label{S18}\\
 &\det(\hat{A}\otimes\hat{D})=[\det(\hat{A})]^{m}[\det(\hat{D})]^{n}\label{S19},
\end{align}
where $\hat{A}$ and $\hat{D}$ are, respectively, $n$- and $m$-square matrices and $\det(\hat{A})$ is different from zero. Then, by applying twice the formula in Eq. \eqref{S18} to $\det[G^{-1}_{\alpha}(i\varepsilon_{n},\mathbf{k})]$, we get
\begin{align}\label{S20}
 \det[G^{-1}_{\alpha}(i\varepsilon_{n},\mathbf{k})]&=\det(-i\varepsilon_{n}+\xi_{p}\tau_{3})\det\left[-i\varepsilon_{n}+\xi_{p}\tau_{3}-\hat{\Gamma}(\mathbf{k})(-i\varepsilon_{n}+\xi_{p}\tau_{3})^{-1}\hat{\Gamma}(\mathbf{k})\right]\det\Bigl\{-i\varepsilon_{n}+\xi_{d}\tau_{3}-i\hat{M}_{\alpha}(\varepsilon_{n},\mathbf{k})\nonumber\\
 &-\hat{\Gamma}^{\dagger}_{x}(k_{x})(-i\varepsilon_{n}+\xi_{p}\tau_{3})^{-1}\hat{\Gamma}_{x}(k_{x})-\Bigl[\hat{\Gamma}^{\dagger}_{y}(k_{y})-\hat{\Gamma}^{\dagger}_{x}(k_{x})(-i\varepsilon_{n}+\xi_{p}\tau_{3})^{-1}\hat{\Gamma}(\mathbf{k})\Bigr]\Bigl[-i\varepsilon_{n}+\xi_{p}\tau_{3}-\hat{\Gamma}(\mathbf{k})\nonumber\\
 &\times(-i\varepsilon_{n}+\xi_{p}\tau_{3})^{-1}\hat{\Gamma}(\mathbf{k})\Bigr]^{-1}\Bigl[\hat{\Gamma}_{y}(k_{y})-\hat{\Gamma}(\mathbf{k})(-i\varepsilon_{n}+\xi_{p}\tau_{3})^{-1}\hat{\Gamma}_{x}(k_{x})\Bigr]\Bigr\},
\end{align}
where $\hat{\Gamma}_{x}(k_{x})=\hat{\Gamma}_{1x}\tau_{3}-\hat{\Gamma}_{2x}k_{x}$, $\hat{\Gamma}_{y}(k_{y})=\hat{\Gamma}_{1y}\tau_{3}-\hat{\Gamma}_{2y}k_{y}$, and $\hat{\Gamma}(\mathbf{k})=\hat{\Gamma}_{1}\tau_{3}-\hat{\Gamma}_{2}(\mathbf{k})$.
The first determinant on the right-hand side of Eq.\eqref{S20} can be easily calculated as
\begin{align}\label{S21}
 \det(-i\varepsilon_{n}+\xi_{p}\tau_{3})&=\det\left[\begin{pmatrix} (-i\varepsilon_{n}+\xi_{p})\mathbb{1}_{L}\otimes\mathbb{1}_{\Lambda}\otimes\mathbb{1}_{\Sigma} & 0 \\ 0 & (-i\varepsilon_{n}-\xi_{p})\mathbb{1}_{L}\otimes\mathbb{1}_{\Lambda}\otimes\mathbb{1}_{\Sigma} \end{pmatrix}_{\tau}\right]\nonumber\\
 &=\det[(-i\varepsilon_{n}+\xi_{p})\mathbb{1}_{L}\otimes\mathbb{1}_{\Lambda}\otimes\mathbb{1}_{\Sigma}]\det[(-i\varepsilon_{n}-\xi_{p})\mathbb{1}_{L}\otimes\mathbb{1}_{\Lambda}\otimes\mathbb{1}_{\Sigma}]\nonumber\\
 &={[\det(-i\varepsilon_{n}+\xi_{p})\mathbb{1}_{L}]}^{4}[\det(\mathbb{1}_{\Lambda}\otimes\mathbb{1}_{\Sigma})]^{2}{\det[(-i\varepsilon_{n}-\xi_{p})\mathbb{1}_{L}]}^{4}[\det(\mathbb{1}_{\Lambda}\otimes\mathbb{1}_{\Sigma})]^{2}\nonumber\\
 &=(\varepsilon^{2}_{n}+\xi_{p}^{2})^{8}.
\end{align}

Before evaluating the second determinant on the right-hand side of Eq. \eqref{S20}, we first write
\begin{align}\label{S22}
 &-i\varepsilon_{n}+\xi_{p}\tau_{3}-\hat{\Gamma}(\mathbf{k})(-i\varepsilon_{n}+\xi_{p}\tau_{3})^{-1}\hat{\Gamma}(\mathbf{k})\nonumber\\
 &=\begin{pmatrix} \begin{pmatrix} \mathcal{G}^{-1}_{L_{x}}(i\varepsilon_{n},\mathbf{k}) & 0 \\ 0 & \mathcal{G}^{-1}_{L_{y}}(i\varepsilon_{n},\mathbf{k}) \end{pmatrix}_{L} & 0 \\ 0 & \Lambda_{1}\begin{pmatrix} [\mathcal{G}^{-1}_{L_{x}}(i\varepsilon_{n},\mathbf{k})]^{\dagger} & 0 \\ 0 & [\mathcal{G}^{-1}_{L_{y}}(i\varepsilon_{n},\mathbf{k})]^{\dagger} \end{pmatrix}_{L}\Lambda_{1} \end{pmatrix}_{\tau}(\tau_{3}\otimes\mathbb{1}_{L}\otimes\mathbb{1}_{\Lambda}\otimes\mathbb{1}_{\Sigma}).
\end{align}
Here, $\Lambda_{1}$ is the Pauli matrix in the pseudospin space $\Lambda$ and the matrices $\mathcal{G}^{-1}_{L_{x}}(i\varepsilon_{n},\mathbf{k})$ and $\mathcal{G}^{-1}_{L_{y}}(i\varepsilon_{n},\mathbf{k})$ appearing above are given by
\begin{align}
 \mathcal{G}^{-1}_{L_{x}}(i\varepsilon_{n},\mathbf{k})=&\begin{pmatrix} -i\varepsilon_{n}+\xi_{p} & 0 \\ 0 & -i\varepsilon_{n}+\xi_{p} \end{pmatrix}_{\Lambda}-t^{2}_{pp}\Bigl(\frac{i\varepsilon_{n}+\xi_{p}}{\varepsilon^{2}_{n}+\xi^{2}_{p}}\Bigr)\begin{pmatrix} a_{1}(\mathbf{k})+b_{1}(\mathbf{k})\Sigma_{3} & 0 \\ 0 & a_{2}(\mathbf{k})+b_{2}(\mathbf{k})\Sigma_{3} \end{pmatrix}_{\Lambda},\label{S23}\\
 \mathcal{G}^{-1}_{L_{y}}(i\varepsilon_{n},\mathbf{k})=&\begin{pmatrix} -i\varepsilon_{n}+\xi_{p} & 0 \\ 0 & -i\varepsilon_{n}+\xi_{p} \end{pmatrix}_{\Lambda}-t^{2}_{pp}\Bigl(\frac{i\varepsilon_{n}+\xi_{p}}{\varepsilon^{2}_{n}+\xi^{2}_{p}}\Bigr)\begin{pmatrix} a_{3}(\mathbf{k})+b_{3}(\mathbf{k})\Sigma_{3} & 0 \\ 0 & a_{4}(\mathbf{k})+b_{4}(\mathbf{k})\Sigma_{3} \end{pmatrix}_{\Lambda},\label{S24}
\end{align}
with $a_{\ell}(\mathbf{k})$ and $b_{\ell}(\mathbf{k})$ $(\ell=1,\ldots,4)$ being functions of the hot spot parameter $\delta=(K_{+}-K_{-})/2$ (see the main text) and the lattice momentum, which we define in Table \ref{TableI}. Thus, we obtain that the determinant of the matrix in Eq. \eqref{S17} evaluates to
\begin{align}\label{S25}
 &\det[-i\varepsilon_{n}+\xi_{p}\tau_{3}-\hat{\Gamma}(\mathbf{k})(-i\varepsilon_{n}+\xi_{p}\tau_{3})^{-1}\hat{\Gamma}(\mathbf{k})]\nonumber\\
 &=\prod^{4}_{\ell=1}\prod_{\sigma=\pm}\left\{\varepsilon^{2}_{n}+\xi^{2}_{p}+2t^{2}_{pp}\left(\frac{\varepsilon^{2}_{n}-\xi^{2}_{p}}{\varepsilon^{2}_{n}+\xi^{2}_{p}}\right)[a_{\ell}(\mathbf{k})+\sigma b_{\ell}(\mathbf{k})]+\frac{t^{4}_{pp}}{\varepsilon^{2}_{n}+\xi^{2}_{p}}[a_{\ell}(\mathbf{k})+\sigma b_{\ell}(\mathbf{k})]^{2}\right\}.
\end{align}

In order to compute the third determinant that appears on the right-hand side of \eqref{S20}, we need to specify the order parameter $\hat{M}_{\alpha}$ of the $d$-wave CDW/PDW order, which will compete with the LC phase. Although the approach developed in this work allows us to address the issue of the interplay of the fully intertwined $d$-wave CDW/PDW order and the LC phase, we will focus, in what follows, only on the cases where there is either a $d$-wave CDW or a $d$-wave PDW competing with the $\Theta_{II}$-loop-current order.

\subsection{Unidirectional $d$-wave CDW (or PDW) order with wavevector $\mathbf{Q}_{x}=(Q_{0},0)$ and LC order}

Before calculating $\det[G^{-1}_{\alpha}(i\varepsilon_{n},\mathbf{k})]$ (for $\alpha=$ CDW-1, PDW-1), we proceed as in Ref. \cite{de_Carvalho} and define the following set of coefficients
\begin{align}
 & c_{1}(k_{x})=\sqrt{2}\left(-t_{pd}+i\frac{R_{II}}{4}k_{x}\right),\label{S26}\\
 & c_{1}(k_{y})=\sqrt{2}\left(-t_{pd}+i\frac{R_{II}}{4}k_{y}\right),\label{S27}\\
 & c_{2}(k_{x})=\frac{\sqrt{2}}{2}(-t_{pd}k_{x}-iR_{II}),\label{S28}\\
 & c_{2}(k_{y})=\frac{\sqrt{2}}{2}(-t_{pd}k_{y}-iR_{II}),\label{S29}
\end{align}
in terms of the Cu-O hopping $t_{pd}$, the LC order parameter $R_{II}$, and the the momentum distance $\mathbf{k}$ to the hot spots. The purpose of defining those four coefficients is to write down $\det[G^{-1}_{\alpha}(i\varepsilon_{n},\mathbf{k})]$ in a compact form. We then construct the set of basis functions shown in Table \ref{TableII} in terms of the hot spot parameter $\delta=(K_{+}-K_{-})/2$ and the $c_{i}(k_{x})$, $c_{i}(k_{y})$ ($i=1,2$).

For a pure unidirectional CDW with a $d$-wave form factor, we should set $\Delta_{\text{PDW}}=0$ in $\hat{M}_{\text{CDW-1/PDW-1}}(i\varepsilon_{n},\mathbf{k})$ (see Eq. \eqref{S13}). As a result, one obtains that the third determinant on the right-hand side of Eq. \eqref{S20} is given by
\begin{align}\label{S30}
 &\det\Bigl\{-i\varepsilon_{n}+\xi_{d}\tau_{3}-i\hat{M}_{\text{CDW-1}}(\varepsilon_{n},\mathbf{k})-\hat{\Gamma}^{\dagger}_{x}(k_{x})(-i\varepsilon_{n}+\xi_{p}\tau_{3})^{-1}\hat{\Gamma}_{x}(k_{x})-\Bigl[\hat{\Gamma}^{\dagger}_{y}(k_{y})-\hat{\Gamma}^{\dagger}_{x}(k_{x})(-i\varepsilon_{n}+\xi_{p}\tau_{3})^{-1}\nonumber\\
 &\times\hat{\Gamma}(\mathbf{k})\Bigr]\Bigl[-i\varepsilon_{n}+\xi_{p}\tau_{3}-\hat{\Gamma}(\mathbf{k})(-i\varepsilon_{n}+\xi_{p}\tau_{3})^{-1}\hat{\Gamma}(\mathbf{k})\Bigr]^{-1}\Bigl[\hat{\Gamma}_{y}(k_{y})-\hat{\Gamma}(\mathbf{k})(-i\varepsilon_{n}+\xi_{p}\tau_{3})^{-1}\hat{\Gamma}_{x}(k_{x})\Bigr]\Bigr\}\nonumber\\
 &=\dfrac{\prod\limits^{2}_{\ell=1}\prod\limits^{2}_{m=1}\mathcal{D}^{\ell m}_{\text{CDW-1}}(i\varepsilon_{n},\mathbf{k})}{(\varepsilon^{2}_{n}+\xi^{2}_{p})^{8}\prod\limits^{4}_{\ell=1}\prod\limits^{}_{\sigma=\pm}\left\{\varepsilon^{2}_{n}+\xi^{2}_{p}+2t^{2}_{pp}\left(\frac{\varepsilon^{2}_{n}-\xi^{2}_{p}}{\varepsilon^{2}_{n}+\xi^{2}_{p}}\right)[a_{\ell}(\mathbf{k})+\sigma b_{\ell}(\mathbf{k})]+\frac{t^{4}_{pp}}{\varepsilon^{2}_{n}+\xi^{2}_{p}}[a_{\ell}(\mathbf{k})+\sigma b_{\ell}(\mathbf{k})]^{2}\right\}}.
\end{align}
The functions $\mathcal{D}^{\ell m}_{\text{CDW-1}}(i\varepsilon_{n},\mathbf{k})$ appearing above are defined as 
\begin{align}
\mathcal{D}^{11}_{\text{CDW-1}}(i\varepsilon_{n},\mathbf{k}) & =|\{(-i\varepsilon_{n}+\xi_{d})[(-i\varepsilon_{n}+\xi_{p})^{2}-t_{pp}^{2}(a_{1}(\mathbf{k})+b_{1}(\mathbf{k}))]-P_{1}^{(0)}(\mathbf{k})(-i\varepsilon_{n}+\xi_{p})-t_{pp}P_{1}^{(1)}(\mathbf{k})\}\nonumber \\
 & \times\{(-i\varepsilon_{n}+\xi_{d})[(-i\varepsilon_{n}+\xi_{p})^{2}-t_{pp}^{2}(a_{3}(\mathbf{k})+b_{3}(\mathbf{k}))]-P_{3}^{(0)}(\mathbf{k})(-i\varepsilon_{n}+\xi_{p})-t_{pp}P_{3}^{(1)}(\mathbf{k})\}\nonumber \\
 & -b^{2}_{\text{CDW-1}}(\varepsilon_{n},\mathbf{k})[(-i\varepsilon_{n}+\xi_{p})^{2}-t_{pp}^{2}(a_{1}(\mathbf{k})+b_{1}(\mathbf{k}))][(-i\varepsilon_{n}+\xi_{p})^{2}-t_{pp}^{2}(a_{3}(\mathbf{k})+b_{3}(\mathbf{k}))]|^{2},\label{S31}\\
\mathcal{D}^{12}_{\text{CDW-1}}(i\varepsilon_{n},\mathbf{k}) & =|\{(-i\varepsilon_{n}+\xi_{d})[(-i\varepsilon_{n}+\xi_{p})^{2}-t_{pp}^{2}(a_{1}(\mathbf{k})-b_{1}(\mathbf{k}))]-M_{1}^{(0)}(\mathbf{k})(-i\varepsilon_{n}+\xi_{p})-t_{pp}M_{1}^{(1)}(\mathbf{k})\}\nonumber \\
 & \times\{(-i\varepsilon_{n}+\xi_{d})[(-i\varepsilon_{n}+\xi_{p})^{2}-t_{pp}^{2}(a_{3}(\mathbf{k})-b_{3}(\mathbf{k}))]-M_{3}^{(0)}(\mathbf{k})(-i\varepsilon_{n}+\xi_{p})-t_{pp}M_{3}^{(1)}(\mathbf{k})\}\nonumber \\
 & -b^{2}_{\text{CDW-1}}(\varepsilon_{n},\mathbf{k})[(-i\varepsilon_{n}+\xi_{p})^{2}-t_{pp}^{2}(a_{1}(\mathbf{k})-b_{1}(\mathbf{k}))][(-i\varepsilon_{n}+\xi_{p})^{2}-t_{pp}^{2}(a_{3}(\mathbf{k})-b_{3}(\mathbf{k}))]|^{2},\label{S32}\\
\mathcal{D}^{21}_{\text{CDW-1}}(i\varepsilon_{n},\mathbf{k}) & =|\{(-i\varepsilon_{n}+\xi_{d})[(-i\varepsilon_{n}+\xi_{p})^{2}-t_{pp}^{2}(a_{2}(\mathbf{k})+b_{2}(\mathbf{k}))]-P_{2}^{(0)}(\mathbf{k})(-i\varepsilon_{n}+\xi_{p})-t_{pp}P_{2}^{(1)}(\mathbf{k})\}\nonumber \\
 & \times\{(-i\varepsilon_{n}+\xi_{d})[(-i\varepsilon_{n}+\xi_{p})^{2}-t_{pp}^{2}(a_{4}(\mathbf{k})+b_{4}(\mathbf{k}))]-P_{4}^{(0)}(\mathbf{k})(-i\varepsilon_{n}+\xi_{p})-t_{pp}P_{4}^{(1)}(\mathbf{k})\}\nonumber \\
 & -b^{2}_{\text{CDW-1}}(\varepsilon_{n},\mathbf{k})[(-i\varepsilon_{n}+\xi_{p})^{2}-t_{pp}^{2}(a_{2}(\mathbf{k})+b_{2}(\mathbf{k}))][(-i\varepsilon_{n}+\xi_{p})^{2}-t_{pp}^{2}(a_{4}(\mathbf{k})+b_{4}(\mathbf{k}))]|^{2},\label{S33}\\
\mathcal{D}^{22}_{\text{CDW-1}}(i\varepsilon_{n},\mathbf{k}) & =|\{(-i\varepsilon_{n}+\xi_{d})[(-i\varepsilon_{n}+\xi_{p})^{2}-t_{pp}^{2}(a_{2}(\mathbf{k})-b_{2}(\mathbf{k}))]-M_{2}^{(0)}(\mathbf{k})(-i\varepsilon_{n}+\xi_{p})-t_{pp}M_{2}^{(1)}(\mathbf{k})\}\nonumber \\
 & \times\{(-i\varepsilon_{n}+\xi_{d})[(-i\varepsilon_{n}+\xi_{p})^{2}-t_{pp}^{2}(a_{4}(\mathbf{k})-b_{4}(\mathbf{k}))]-M_{4}^{(0)}(\mathbf{k})(-i\varepsilon_{n}+\xi_{p})-t_{pp}M_{4}^{(1)}(\mathbf{k})\}\nonumber \\
 & -b^{2}_{\text{CDW-1}}(\varepsilon_{n},\mathbf{k})[(-i\varepsilon_{n}+\xi_{p})^{2}-t_{pp}^{2}(a_{2}(\mathbf{k})-b_{2}(\mathbf{k}))][(-i\varepsilon_{n}+\xi_{p})^{2}-t_{pp}^{2}(a_{4}(\mathbf{k})-b_{4}(\mathbf{k}))]|^{2}.\label{S34}
\end{align}

\begin{table}
\begin{ruledtabular}
\centering
\caption{First set of basis functions needed to compute the functional free energy of the present three-band model. In our notation, the indices $\ell$ and $\widetilde{\ell}$ refer, respectively, to the functions $a_{\ell}(\mathbf{k})$ [and $b_{\ell}(\mathbf{k})$] and $\widetilde{a}_{\ell}(\mathbf{k})$ [and $\widetilde{b}_{\ell}(\mathbf{k})$].}\label{TableI}
\begin{tabular}{ccccc}
$\ell$         & \ & $a_{\ell}(\mathbf{k})$ & \ & $b_{\ell}(\mathbf{k})$ \\ \hline
$1$  & \ &  $(k_{x}+k_{y})^{2}+\sin^{2}\delta(k_{y}-k_{x}+2\cot\delta)^{2}$  & \ &  $2\sin\delta[(k_{y}+\cot\delta)^{2}-(k_{x}-\cot\delta)^{2}]$   \\
$2$  & \ &  $(k_{x}+k_{y})^{2}+\sin^{2}\delta(k_{y}-k_{x}-2\cot\delta)^{2}$  & \ &  $2\sin\delta[(k_{y}-\cot\delta)^{2}-(k_{x}+\cot\delta)^{2}]$   \\
$3$  & \ &  $(k_{x}-k_{y})^{2}+\sin^{2}\delta(k_{x}+k_{y}+2\cot\delta)^{2}$  & \ &  $2\sin\delta[(k_{x}+\cot\delta)^{2}-(k_{y}+\cot\delta)^{2}]$   \\
$4$  & \ &  $(k_{x}-k_{y})^{2}+\sin^{2}\delta(k_{x}+k_{y}-2\cot\delta)^{2}$  & \ &  $2\sin\delta[(k_{x}-\cot\delta)^{2}-(k_{y}-\cot\delta)^{2}]$   \\
$\widetilde{1}$  & \ &  $-\sin\delta(k_{x}-k_{y}-2\cot\delta)$  & \ &  $k_{x}+k_{y}$   \\
$\widetilde{2}$  & \ &  $\sin\delta(k_{x}-k_{y}+2\cot\delta)$  & \ &  $-(k_{x}+k_{y})$   \\
$\widetilde{3}$  & \ &  $\sin\delta(k_{x}+k_{y}+2\cot\delta)$  & \ &  $k_{x}-k_{y}$   \\
$\widetilde{4}$  & \ &  $-\sin\delta(k_{x}+k_{y}-2\cot\delta)$  & \ &  $-(k_{x}-k_{y})$   \\
\end{tabular}
\end{ruledtabular}
\end{table}

\noindent Here, we have used the set of basis functions given in Table \ref{TableI}, as well as the new functions
\begin{align}
P_{\ell}^{(0)}(\mathbf{k}) & =a_{\ell x}(k_{x})+a_{\ell y}(k_{y})+b_{\ell x}(k_{x})+b_{\ell y}(k_{y}),\label{S35}\\
P_{\ell}^{(1)}(\mathbf{k}) & =[\widetilde{a}_{\ell}(\mathbf{k})+\widetilde{b}_{\ell}(\mathbf{k})][a_{\ell xy}(\mathbf{k})+b_{\ell xy}(\mathbf{k})],\label{S36}\\
M_{\ell}^{(0)}(\mathbf{k}) & =a_{\ell x}(k_{x})+a_{\ell y}(k_{y})-b_{\ell x}(k_{x})-b_{\ell y}(k_{y}),\label{S37}\\
M_{\ell}^{(1)}(\mathbf{k}) & =[\widetilde{a}_{\ell}(\mathbf{k})-\widetilde{b}_{\ell}(\mathbf{k})][a_{\ell xy}(\mathbf{k})-b_{\ell xy}(\mathbf{k})],\label{S38}
\end{align}
which have in their definition functions of both Tables \ref{TableI} and \ref{TableII}.

\begin{table}
\begin{ruledtabular}
\centering
\caption{Second set of basis functions used to represent the functional free energy $\mathcal{F}_{\alpha}[T,n_{p},R_{II},\hat{M}_{\alpha}]$. Here, these functions are written in terms of the hot spot parameter $\delta=(K_{+}-K_{-})/2$ and the coefficients $c_{i}(k_{x})$ and $c_{i}(k_{y})$ ($i=1,2$) \cite{de_Carvalho}.}\label{TableII}
\begin{tabular}{lc}
Basis function         & Definition \\ \hline
$a_{1x}(k_{x})$  &  $|c_{1}(k_{x})|^{2}+|c_{2}(k_{x})|^{2}$   \\
$a_{2x}(k_{x})$  &  $|c_{1}(k_{x})|^{2}+|c_{2}(k_{x})|^{2}$   \\
$a_{3x}(k_{x})$  &  $|c_{1}(k_{x})|^{2}+|c_{2}(k_{x})|^{2}$   \\
$a_{4x}(k_{x})$  &  $|c_{1}(k_{x})|^{2}+|c_{2}(k_{x})|^{2}$   \\
$b_{1x}(k_{x})$  &  $\sin\delta[|c_{1}(k_{x})|^{2}-|c_{2}(k_{x})|^{2}]+2\cos\delta\operatorname{Re}[c^{*}_{1}(k_{x})c_{2}(k_{x})]$   \\
$b_{2x}(k_{x})$  &  $\sin\delta[|c_{1}(k_{x})|^{2}-|c_{2}(k_{x})|^{2}]-2\cos\delta\operatorname{Re}[c^{*}_{1}(k_{x})c_{2}(k_{x})]$   \\
$b_{3x}(k_{x})$  &  $\sin\delta[|c_{2}(k_{x})|^{2}-|c_{1}(k_{x})|^{2}]+2\cos\delta\operatorname{Re}[c^{*}_{1}(k_{x})c_{2}(k_{x})]$   \\
$b_{4x}(k_{x})$  &  $\sin\delta[|c_{2}(k_{x})|^{2}-|c_{1}(k_{x})|^{2}]-2\cos\delta\operatorname{Re}[c^{*}_{1}(k_{x})c_{2}(k_{x})]$   \\
$a_{1y}(k_{y})$  &  $|c_{1}(k_{y})|^{2}+|c_{2}(k_{y})|^{2}$   \\
$a_{2y}(k_{y})$  &  $|c_{1}(k_{y})|^{2}+|c_{2}(k_{y})|^{2}$   \\
$a_{3y}(k_{y})$  &  $|c_{1}(k_{y})|^{2}+|c_{2}(k_{y})|^{2}$   \\
$a_{4y}(k_{y})$  &  $|c_{1}(k_{y})|^{2}+|c_{2}(k_{y})|^{2}$   \\
$b_{1y}(k_{y})$  &  $\sin\delta[|c_{2}(k_{y})|^{2}-|c_{1}(k_{y})|^{2}]+2\cos\delta\operatorname{Re}[c^{*}_{1}(k_{y})c_{2}(k_{y})]$   \\
$b_{2y}(k_{y})$  &  $\sin\delta[|c_{2}(k_{y})|^{2}-|c_{1}(k_{y})|^{2}]-2\cos\delta\operatorname{Re}[c^{*}_{1}(k_{y})c_{2}(k_{y})]$   \\
$b_{3y}(k_{y})$  &  $\sin\delta[|c_{1}(k_{y})|^{2}-|c_{2}(k_{y})|^{2}]-2\cos\delta\operatorname{Re}[c^{*}_{1}(k_{y})c_{2}(k_{y})]$   \\
$b_{4y}(k_{y})$  &  $\sin\delta[|c_{1}(k_{y})|^{2}-|c_{2}(k_{y})|^{2}]+2\cos\delta\operatorname{Re}[c^{*}_{1}(k_{y})c_{2}(k_{y})]$   \\
$a_{1xy}(\mathbf{k})$  &  $2\cos\delta\operatorname{Re}[c^{*}_{1}(k_{x})c_{1}(k_{y})+c^{*}_{2}(k_{x})c_{2}(k_{y})]+2\sin\delta\operatorname{Re}[c^{*}_{1}(k_{x})c_{2}(k_{y})-c^{*}_{2}(k_{x})c_{1}(k_{y})]$   \\
$a_{2xy}(\mathbf{k})$  &  $2\cos\delta\operatorname{Re}[c^{*}_{1}(k_{x})c_{1}(k_{y})+c^{*}_{2}(k_{x})c_{2}(k_{y})]-2\sin\delta\operatorname{Re}[c^{*}_{1}(k_{x})c_{2}(k_{y})-c^{*}_{2}(k_{x})c_{1}(k_{y})]$   \\
$a_{3xy}(\mathbf{k})$  &  $2\cos\delta\operatorname{Re}[c^{*}_{1}(k_{x})c_{1}(k_{y})-c^{*}_{2}(k_{x})c_{2}(k_{y})]+2\sin\delta\operatorname{Re}[c^{*}_{1}(k_{x})c_{2}(k_{y})+c^{*}_{2}(k_{x})c_{1}(k_{y})]$   \\
$a_{4xy}(\mathbf{k})$  &  $2\cos\delta\operatorname{Re}[c^{*}_{1}(k_{x})c_{1}(k_{y})-c^{*}_{2}(k_{x})c_{2}(k_{y})]-2\sin\delta\operatorname{Re}[c^{*}_{1}(k_{x})c_{2}(k_{y})+c^{*}_{2}(k_{x})c_{1}(k_{y})]$   \\
$b_{1xy}(\mathbf{k})$  &  $2\operatorname{Re}[c^{*}_{2}(k_{x})c_{1}(k_{y})+c^{*}_{1}(k_{x})c_{2}(k_{y})]$   \\
$b_{2xy}(\mathbf{k})$  & $-2\operatorname{Re}[c^{*}_{2}(k_{x})c_{1}(k_{y})+c^{*}_{1}(k_{x})c_{2}(k_{y})]$    \\
$b_{3xy}(\mathbf{k})$  &  $2\operatorname{Re}[c^{*}_{2}(k_{x})c_{1}(k_{y})-c^{*}_{1}(k_{x})c_{2}(k_{y})]$   \\
$b_{4xy}(\mathbf{k})$  &  $-2\operatorname{Re}[c^{*}_{2}(k_{x})c_{1}(k_{y})-c^{*}_{1}(k_{x})c_{2}(k_{y})]$   \\
\end{tabular}
\end{ruledtabular}
\end{table}

From the results above, we conclude that the determinant in \eqref{S20} can be written as
\begin{equation}\label{S39}
\det[G^{-1}_{\text{CDW-1}}(i\varepsilon_{n},\mathbf{k})]=\prod\limits _{\ell=1}^{2}\prod\limits _{m=1}^{2}\mathcal{D}^{\ell m}_{\text{CDW-1}}(i\varepsilon_{n},\mathbf{k}).
\end{equation}
Then, the mean-field equations describing the interplay between unidirectional CDW with a $d$-wave form factor and the $\Theta_{II}$-loop current order are simply given by
\begin{align}
 b_{\text{CDW-1}}(\varepsilon_{n},\mathbf{k})&=\frac{3\lambda^{2}T}{16}\sum\limits^{2}_{\ell=1}\sum\limits^{2}_{m=1}\sum\limits^{}_{\varepsilon'_{n}}\int\dfrac{D_{eff}(\varepsilon_{n}-\varepsilon'_{n},\mathbf{k}-\mathbf{k}')}{\mathcal{D}^{\ell m}_{\text{CDW-1}}(i\varepsilon'_{n},\mathbf{k}')}\dfrac{\partial\mathcal{D}^{\ell m}_{\text{CDW-1}}(i\varepsilon'_{n},\mathbf{k}')}{\partial b_{\text{CDW-1}}(\varepsilon'_{n},\mathbf{k}')}\frac{d\mathbf{k}'}{(2\pi)^{2}},\label{S40}\\
 R_{II}&=\frac{V_{pd}T}{2}\sum\limits^{2}_{\ell=1}\sum\limits^{2}_{m=1}\sum_{\varepsilon_{n}}\int\frac{1}{\mathcal{D}^{\ell m}_{\text{CDW-1}}(i\varepsilon_{n},\mathbf{k})}\frac{\partial\mathcal{D}^{\ell m}_{\text{CDW-1}}(i\varepsilon_{n},\mathbf{k})}{\partial R_{II}}\frac{d\mathbf{k}}{(2\pi)^{2}}.\label{S41}
\end{align}
On the other hand, when unidirectional $d$-wave CDW is substituted by its SU(2) partner (i.e. the unidirectional $d$-wave PDW) in the above analysis, the mean-field equations describing such a competition with the LC order become 
\begin{align}
 b_{\text{PDW-1}}(\varepsilon_{n},\mathbf{k})&=\frac{3\lambda^{2}T}{16}\sum\limits^{2}_{\ell=1}\sum\limits^{2}_{m=1}\sum\limits^{}_{\varepsilon'_{n}}\int\dfrac{D_{eff}(\varepsilon_{n}-\varepsilon'_{n},\mathbf{k}-\mathbf{k}')}{\mathcal{D}^{\ell m}_{\text{PDW-1}}(i\varepsilon'_{n},\mathbf{k}')}\dfrac{\partial\mathcal{D}^{\ell m}_{\text{PDW-1}}(i\varepsilon'_{n},\mathbf{k}')}{\partial b_{\text{PDW-1}}(\varepsilon'_{n},\mathbf{k}')}\frac{d\mathbf{k}'}{(2\pi)^{2}},\label{S42}\\
 R_{II}&=\frac{V_{pd}T}{2}\sum\limits^{2}_{\ell=1}\sum\limits^{2}_{m=1}\sum_{\varepsilon_{n}}\int\frac{1}{\mathcal{D}^{\ell m}_{\text{PDW-1}}(i\varepsilon_{n},\mathbf{k})}\frac{\partial\mathcal{D}^{\ell m}_{\text{PDW-1}}(i\varepsilon_{n},\mathbf{k})}{\partial R_{II}}\frac{d\mathbf{k}}{(2\pi)^{2}},\label{S43}
\end{align}
where the new functions $\mathcal{D}^{\ell m}_{\text{PDW-1}}(i\varepsilon_{n},\mathbf{k})$ are defined as
\begin{align}
\mathcal{D}^{11}_{\text{PDW-1}}(i\varepsilon_{n},\mathbf{k}) & =|\{(i\varepsilon_{n}+\xi_{d})[(i\varepsilon_{n}+\xi_{p})^{2}-t_{pp}^{2}(a_{2}(\mathbf{k})+b_{2}(\mathbf{k}))]-P_{2}^{(0)}(\mathbf{k})(i\varepsilon_{n}+\xi_{p})-t_{pp}P_{2}^{(1)}(\mathbf{k})\}\nonumber \\
 & \times\{(-i\varepsilon_{n}+\xi_{d})[(-i\varepsilon_{n}+\xi_{p})^{2}-t_{pp}^{2}(a_{3}(\mathbf{k})+b_{3}(\mathbf{k}))]-P_{3}^{(0)}(\mathbf{k})(-i\varepsilon_{n}+\xi_{p})-t_{pp}P_{3}^{(1)}(\mathbf{k})\}\nonumber \\
 & -b^{2}_{\text{PDW-1}}(\varepsilon_{n},\mathbf{k})[(i\varepsilon_{n}+\xi_{p})^{2}-t_{pp}^{2}(a_{2}(\mathbf{k})+b_{2}(\mathbf{k}))][(-i\varepsilon_{n}+\xi_{p})^{2}-t_{pp}^{2}(a_{3}(\mathbf{k})+b_{3}(\mathbf{k}))]|^{2},\label{S44}\\
\mathcal{D}^{12}_{\text{PDW-1}}(i\varepsilon_{n},\mathbf{k}) & =|\{(i\varepsilon_{n}+\xi_{d})[(i\varepsilon_{n}+\xi_{p})^{2}-t_{pp}^{2}(a_{2}(\mathbf{k})-b_{2}(\mathbf{k}))]-M_{2}^{(0)}(\mathbf{k})(i\varepsilon_{n}+\xi_{p})-t_{pp}M_{2}^{(1)}(\mathbf{k})\}\nonumber \\
 & \times\{(-i\varepsilon_{n}+\xi_{d})[(-i\varepsilon_{n}+\xi_{p})^{2}-t_{pp}^{2}(a_{3}(\mathbf{k})-b_{3}(\mathbf{k}))]-M_{3}^{(0)}(\mathbf{k})(-i\varepsilon_{n}+\xi_{p})-t_{pp}M_{3}^{(1)}(\mathbf{k})\}\nonumber \\
 & -b^{2}_{\text{PDW-1}}(\varepsilon_{n},\mathbf{k})[(i\varepsilon_{n}+\xi_{p})^{2}-t_{pp}^{2}(a_{2}(\mathbf{k})-b_{2}(\mathbf{k}))][(-i\varepsilon_{n}+\xi_{p})^{2}-t_{pp}^{2}(a_{3}(\mathbf{k})-b_{3}(\mathbf{k}))]|^{2},\label{S45}\\
\mathcal{D}^{21}_{\text{PDW-1}}(i\varepsilon_{n},\mathbf{k}) & =|\{(i\varepsilon_{n}+\xi_{d})[(i\varepsilon_{n}+\xi_{p})^{2}-t_{pp}^{2}(a_{1}(\mathbf{k})+b_{1}(\mathbf{k}))]-P_{1}^{(0)}(\mathbf{k})(i\varepsilon_{n}+\xi_{p})-t_{pp}P_{1}^{(1)}(\mathbf{k})\}\nonumber \\
 & \times\{(-i\varepsilon_{n}+\xi_{d})[(-i\varepsilon_{n}+\xi_{p})^{2}-t_{pp}^{2}(a_{4}(\mathbf{k})+b_{4}(\mathbf{k}))]-P_{4}^{(0)}(\mathbf{k})(-i\varepsilon_{n}+\xi_{p})-t_{pp}P_{4}^{(1)}(\mathbf{k})\}\nonumber \\
 & -b^{2}_{\text{PDW-1}}(\varepsilon_{n},\mathbf{k})[(i\varepsilon_{n}+\xi_{p})^{2}-t_{pp}^{2}(a_{1}(\mathbf{k})+b_{1}(\mathbf{k}))][(-i\varepsilon_{n}+\xi_{p})^{2}-t_{pp}^{2}(a_{4}(\mathbf{k})+b_{4}(\mathbf{k}))]|^{2},\label{S46}\\
\mathcal{D}^{22}_{\text{PDW-1}}(i\varepsilon_{n},\mathbf{k}) & =|\{(i\varepsilon_{n}+\xi_{d})[(i\varepsilon_{n}+\xi_{p})^{2}-t_{pp}^{2}(a_{1}(\mathbf{k})-b_{1}(\mathbf{k}))]-M_{1}^{(0)}(\mathbf{k})(i\varepsilon_{n}+\xi_{p})-t_{pp}M_{1}^{(1)}(\mathbf{k})\}\nonumber \\
 & \times\{(-i\varepsilon_{n}+\xi_{d})[(-i\varepsilon_{n}+\xi_{p})^{2}-t_{pp}^{2}(a_{4}(\mathbf{k})-b_{4}(\mathbf{k}))]-M_{4}^{(0)}(\mathbf{k})(-i\varepsilon_{n}+\xi_{p})-t_{pp}M_{4}^{(1)}(\mathbf{k})\}\nonumber \\
 & -b^{2}_{\text{PDW-1}}(\varepsilon_{n},\mathbf{k})[(i\varepsilon_{n}+\xi_{p})^{2}-t_{pp}^{2}(a_{1}(\mathbf{k})-b_{1}(\mathbf{k}))][(-i\varepsilon_{n}+\xi_{p})^{2}-t_{pp}^{2}(a_{4}(\mathbf{k})-b_{4}(\mathbf{k}))]|^{2}.\label{S47}
\end{align}

In order to solve those mean-field equations, we first have to evaluate the Matsubara sums appearing in them. To do this analytically, we follow Ref. \cite{de_Carvalho} and neglect the frequency-momentum dependency of $b_{\alpha}(\varepsilon_{n},\mathbf{k})$. As a consequence, the Matsubara sums are transformed into integrals by means of the residue theorem when one performs the analytic continuation $\varepsilon_{n}\rightarrow-iz$ for each $\mathcal{D}^{\ell m}_{\alpha}(i\varepsilon_{n},\mathbf{k})$. It turns out that by observing the definition of both $\mathcal{D}^{\ell m}_{\text{CDW-1}}(i\varepsilon_{n},\mathbf{k})$ and $\mathcal{D}^{\ell m}_{\text{PDW-1}}(i\varepsilon_{n},\mathbf{k})$, we conclude that each $\mathcal{D}^{\ell m}_{\alpha}(i\varepsilon_{n},\mathbf{k})$ may be written as a product of a function $h^{\ell m}_{\alpha}(i\varepsilon_{n},\mathbf{k})$ times its complex conjugate $\xoverline[0.8]{h}^{\ell m}_{\alpha}(i\varepsilon_{n},\mathbf{k})$. Therefore, we can make the analytic continuation 
for 
$\mathcal{D}^{\ell m}_{\alpha}(i\varepsilon_{n},\mathbf{k})$ as follows 
\begin{equation}\label{S48}
\mathcal{D}^{\ell m}_{\alpha}(z,\mathbf{k})=h^{\ell m}_{\alpha}(z,\mathbf{k})\xoverline[0.8]{h}^{\ell m}_{\alpha}(z,\mathbf{k}).
\end{equation}
By determining the roots of $h^{\ell m}_{\alpha}(z,\mathbf{k})$ and $\xoverline[0.8]{h}^{\ell m}_{\alpha}(z,\mathbf{k})$ with respect to the variable $z$, all complex integrals for each mean-field equation can be evaluated with an appropriate contour of integration. Then, we solve those mean-field equations, as well as the remaining momentum integrals numerically.

\subsection{Bidirectional $d$-wave CDW (or PDW) order with wavevectors $\mathbf{Q}_{x}=(Q_{0},0)$ and $\mathbf{Q}_{y}=(0,Q_{0})$ and LC order}

In order to study the mutual interplay of bidirectional CDW (or PDW) with a $d$-wave form factor and the LC order, we follow a similar procedure already explained in the last section for the case of unidirectional $d$-wave CDW/PDW competing with the LC order parameter. As a result, the mean-field equations for the bidirectional case can be written as
\begin{align}
 b_{\alpha}(\varepsilon_{n},\mathbf{k})&=\frac{3\lambda^{2}T}{32}\sum\limits^{1}_{\ell=1}\sum\limits^{2}_{m=1}\sum\limits^{}_{\varepsilon'_{n}}\int\dfrac{D_{eff}(\varepsilon_{n}-\varepsilon'_{n},\mathbf{k}-\mathbf{k}')}{\mathcal{D}^{\ell m}_{\alpha}(i\varepsilon'_{n},\mathbf{k}')}\dfrac{\partial\mathcal{D}^{\ell m}_{\alpha}(i\varepsilon'_{n},\mathbf{k}')}{\partial b_{\alpha}(\varepsilon'_{n},\mathbf{k}')}\frac{d\mathbf{k}'}{(2\pi)^{2}},\label{S49}\\
 R_{II}&=\frac{V_{pd}T}{2}\sum\limits^{1}_{\ell=1}\sum\limits^{2}_{m=1}\sum_{\varepsilon_{n}}\int\frac{1}{\mathcal{D}^{\ell m}_{\alpha}(i\varepsilon_{n},\mathbf{k})}\frac{\partial\mathcal{D}^{\ell m}_{\alpha}(i\varepsilon_{n},\mathbf{k})}{\partial R_{II}}\frac{d\mathbf{k}}{(2\pi)^{2}},\label{S50}
\end{align}
with the indices in the subscript of the order parameters standing for $\alpha=$ CDW-2, PDW-2. Here, the polynomial complex functions $\mathcal{D}^{\ell m}_{\text{CDW-2}}(i\varepsilon_{n},\mathbf{k})$ have the form
\begin{align}
\mathcal{D}^{11}_{\text{CDW-2}}(i\varepsilon_{n},\mathbf{k}) & =|\{(-i\varepsilon_{n}+\xi_{d})[(-i\varepsilon_{n}+\xi_{p})^{2}-t_{pp}^{2}(a_{1}(\mathbf{k})+b_{1}(\mathbf{k}))]-P_{1}^{(0)}(\mathbf{k})(-i\varepsilon_{n}+\xi_{p})-t_{pp}P_{1}^{(1)}(\mathbf{k})\}\nonumber \\
 & \times\{(-i\varepsilon_{n}+\xi_{d})[(-i\varepsilon_{n}+\xi_{p})^{2}-t_{pp}^{2}(a_{2}(\mathbf{k})+b_{2}(\mathbf{k}))]-P_{2}^{(0)}(\mathbf{k})(-i\varepsilon_{n}+\xi_{p})-t_{pp}P_{2}^{(1)}(\mathbf{k})\}\nonumber \\
 & \times\{(-i\varepsilon_{n}+\xi_{d})[(-i\varepsilon_{n}+\xi_{p})^{2}-t_{pp}^{2}(a_{3}(\mathbf{k})+b_{3}(\mathbf{k}))]-P_{3}^{(0)}(\mathbf{k})(-i\varepsilon_{n}+\xi_{p})-t_{pp}P_{3}^{(1)}(\mathbf{k})\}\nonumber \\
 & \times\{(-i\varepsilon_{n}+\xi_{d})[(-i\varepsilon_{n}+\xi_{p})^{2}-t_{pp}^{2}(a_{4}(\mathbf{k})+b_{4}(\mathbf{k}))]-P_{4}^{(0)}(\mathbf{k})(-i\varepsilon_{n}+\xi_{p})-t_{pp}P_{4}^{(1)}(\mathbf{k})\}\nonumber \\
 & -b^{2}_{\text{CDW-2}}(\varepsilon_{n},\mathbf{k})\{(-i\varepsilon_{n}+\xi_{d})[(-i\varepsilon_{n}+\xi_{p})^{2}-t_{pp}^{2}(a_{1}(\mathbf{k})+b_{1}(\mathbf{k}))][(-i\varepsilon_{n}+\xi_{p})^{2}-t_{pp}^{2}(a_{2}(\mathbf{k})\nonumber\\
 & +b_{2}(\mathbf{k}))]-[P_{1}^{(0)}(\mathbf{k})(-i\varepsilon_{n}+\xi_{p})+t_{pp}P_{1}^{(1)}(\mathbf{k})][(-i\varepsilon_{n}+\xi_{p})^{2}-t_{pp}^{2}(a_{2}(\mathbf{k})+b_{2}(\mathbf{k}))]-[P_{2}^{(0)}(\mathbf{k})\nonumber\\
 & \times(-i\varepsilon_{n}+\xi_{p})+t_{pp}P_{2}^{(1)}(\mathbf{k})][(-i\varepsilon_{n}+\xi_{p})^{2}-t_{pp}^{2}(a_{1}(\mathbf{k})+b_{1}(\mathbf{k}))]\}\{(-i\varepsilon_{n}+\xi_{d})[(-i\varepsilon_{n}+\xi_{p})^{2}\nonumber\\
 & -t_{pp}^{2}(a_{3}(\mathbf{k})+b_{3}(\mathbf{k}))][(-i\varepsilon_{n}+\xi_{p})^{2}-t_{pp}^{2}(a_{4}(\mathbf{k})+b_{4}(\mathbf{k}))]-[P_{3}^{(0)}(\mathbf{k})(-i\varepsilon_{n}+\xi_{p})+t_{pp}P_{3}^{(1)}(\mathbf{k})]\nonumber\\
 &\times [(-i\varepsilon_{n}+\xi_{p})^{2}-t_{pp}^{2}(a_{4}(\mathbf{k})+b_{4}(\mathbf{k}))]-[P_{4}^{(0)}(\mathbf{k})(-i\varepsilon_{n}+\xi_{p})+t_{pp}P_{4}^{(1)}(\mathbf{k})][(-i\varepsilon_{n}+\xi_{p})^{2}\nonumber\\
 & -t_{pp}^{2}(a_{3}(\mathbf{k})+b_{3}(\mathbf{k}))]\}|^{2},\label{S51}
 \end{align}

 \begin{align}
 \mathcal{D}^{12}_{\text{CDW-2}}(i\varepsilon_{n},\mathbf{k}) & =|\{(-i\varepsilon_{n}+\xi_{d})[(-i\varepsilon_{n}+\xi_{p})^{2}-t_{pp}^{2}(a_{1}(\mathbf{k})-b_{1}(\mathbf{k}))]-M_{1}^{(0)}(\mathbf{k})(-i\varepsilon_{n}+\xi_{p})-t_{pp}M_{1}^{(1)}(\mathbf{k})\}\nonumber\\ 
 & \times\{(-i\varepsilon_{n}+\xi_{d})[(-i\varepsilon_{n}+\xi_{p})^{2}-t_{pp}^{2}(a_{2}(\mathbf{k})-b_{2}(\mathbf{k}))]-M_{2}^{(0)}(\mathbf{k})(-i\varepsilon_{n}+\xi_{p})-t_{pp}M_{2}^{(1)}(\mathbf{k})\}\nonumber \\
 & \times\{(-i\varepsilon_{n}+\xi_{d})[(-i\varepsilon_{n}+\xi_{p})^{2}-t_{pp}^{2}(a_{3}(\mathbf{k})-b_{3}(\mathbf{k}))]-M_{3}^{(0)}(\mathbf{k})(-i\varepsilon_{n}+\xi_{p})-t_{pp}M_{3}^{(1)}(\mathbf{k})\}\nonumber \\
 & \times\{(-i\varepsilon_{n}+\xi_{d})[(-i\varepsilon_{n}+\xi_{p})^{2}-t_{pp}^{2}(a_{4}(\mathbf{k})-b_{4}(\mathbf{k}))]-M_{4}^{(0)}(\mathbf{k})(-i\varepsilon_{n}+\xi_{p})-t_{pp}M_{4}^{(1)}(\mathbf{k})\}\nonumber \\
 & -b^{2}_{\text{CDW-2}}(\varepsilon_{n},\mathbf{k})\{(-i\varepsilon_{n}+\xi_{d})[(-i\varepsilon_{n}+\xi_{p})^{2}-t_{pp}^{2}(a_{1}(\mathbf{k})-b_{1}(\mathbf{k}))][(-i\varepsilon_{n}+\xi_{p})^{2}-t_{pp}^{2}(a_{2}(\mathbf{k})\nonumber\\
 & -b_{2}(\mathbf{k}))]-[M_{1}^{(0)}(\mathbf{k})(-i\varepsilon_{n}+\xi_{p})+t_{pp}M_{1}^{(1)}(\mathbf{k})][(-i\varepsilon_{n}+\xi_{p})^{2}-t_{pp}^{2}(a_{2}(\mathbf{k})-b_{2}(\mathbf{k}))]-[M_{2}^{(0)}(\mathbf{k})\nonumber\\
 & \times(-i\varepsilon_{n}+\xi_{p})+t_{pp}M_{2}^{(1)}(\mathbf{k})][(-i\varepsilon_{n}+\xi_{p})^{2}-t_{pp}^{2}(a_{1}(\mathbf{k})-b_{1}(\mathbf{k}))]\}\{(-i\varepsilon_{n}+\xi_{d})[(-i\varepsilon_{n}+\xi_{p})^{2}\nonumber\\
 & -t_{pp}^{2}(a_{3}(\mathbf{k})-b_{3}(\mathbf{k}))][(-i\varepsilon_{n}+\xi_{p})^{2}-t_{pp}^{2}(a_{4}(\mathbf{k})-b_{4}(\mathbf{k}))]-[M_{3}^{(0)}(\mathbf{k})(-i\varepsilon_{n}+\xi_{p})+t_{pp}M_{3}^{(1)}(\mathbf{k})]\nonumber\\
 &\times [(-i\varepsilon_{n}+\xi_{p})^{2}-t_{pp}^{2}(a_{4}(\mathbf{k})-b_{4}(\mathbf{k}))]-[M_{4}^{(0)}(\mathbf{k})(-i\varepsilon_{n}+\xi_{p})+t_{pp}M_{4}^{(1)}(\mathbf{k})][(-i\varepsilon_{n}+\xi_{p})^{2}\nonumber\\
 & -t_{pp}^{2}(a_{3}(\mathbf{k})-b_{3}(\mathbf{k}))]\}|^{2},\label{S52}
\end{align}
with similar functions for PDW-2. The method for solving Eqs. \eqref{S49} and \eqref{S50} is identical to the one already described in the last section. The corresponding results are discussed in the main text of this paper.


%

\end{document}